\documentclass{elsart}
\newcommand{\bea}{\begin{eqnarray}}
\newcommand{\eea}{\end{eqnarray}}
\newcommand{\rvec}{\vec{r}}

\newcommand{\eg}{\textit{e.g.}}
\newcommand{\ie}{\textit{i.e.}}

\usepackage{graphics}
\usepackage{color} %monochrome,usenames
\usepackage[normalem]{ulem}
\usepackage{colortbl}

\begin{document}
%\runauthor{Winston Pooh)
\begin{frontmatter}
\title{Initiating a Mexican wave: An instantaneous collective 
decision with both short and long range interactions}
\author{Ill\'es J. Farkas and Tam\'as Vicsek}

%\address{$^1$Biological Physics Research Group of HAS,
%P\'azm\'any P.\ stny.\ 1A, H-1117 Budapest, Hungary}

\address{Biological Physics Research Group of HAS and
Department of Biological Physics, E\"otv\"os University,
P\'azm\'any P.\ stny.\ 1A, H-1117 Budapest, Hungary.}

\begin{abstract}
An interesting example for
collective decision making is 
the so-called Mexican wave during which the
spectators in a 
stadium leap to their feet with their arms up and then sit down again
following those to their left (right) with a small delay. 
Here we use a simple, but
realistic model 
to explain how the combination of the local and global interactions of
the spectators produces a breaking of the symmetry 
resulting in the replacement of the 
symmetric solution 
-- containing two propagating waves -- 
by a single wave moving in one of 
the two possible directions. Our model is based on and compared to the
extensive observations of volunteers filling out the 
related questionnaire we have posted on the Internet. We find that, as
a function of the parameter controlling the strength 
of the global interactions, the transition to the single wave solution
has features reminiscent of discontinuous 
transitions. After
the spontaneous symmetry breaking the two directions of propagation are
still statistically equivalent. We investigate also how this remaining
symmetry is broken in real stadia by a small asymmetrical term in the
perception of spectators.


% 05.45.Yv      Solitons
% 05.65.+b      Self-organized systems
% 87.23.Ge      Dynamics of social systems
% 89.65.-s      Social and economic systems
% 89.75.Kd      Patterns
%
%
% 47.54.+r Pattern selection; pattern formation

\end{abstract}

\end{frontmatter}

\section{Introduction}

In recent years,
the rapid development 
of observational methods
and computational power has made it possible
to accumulate data on the {\it collective motion}
of a large number of living organisms
\cite{parrish}
and to investigate the few observed universal 
patterns of motion
also by computational tools and  
{\it statistical physics based models}
\cite{SchPED,novel,annPhys,panic}.
Similarly to collective motion,
the number of participants in
{\it collective opinion formation and decisions} 
is often very large;
a key factor is 
interaction (influence and imitation)
between the participants, which strongly
reduces the number of possible global patterns 
(see, \eg, Ref.~\cite{turner}) suggesting
that the number of relevant parameters is small.
Statistical physics based models have been successfully applied in
the analysis of collective opinion formation and decisions 
%as well \cite{turner,galam,sznajd,fortunato,szabo,bouchaud,schweitzer}.
as well. 
Unanimous and undecided election results and cooperation phenomena
have been described by models containing
particles with
a small number of allowed states
and simple rules of interaction plus external fields
\cite{galam,sznajd,fortunato,szabo}.
Surprisingly, 
within the same modelling framework one can 
explain the shape of the transition measured for the cumulated
binary decisions of millions of humans
in several further collective decision processes 
of high public interest, such as birth rate and cell phone 
purchases \cite{bouchaud}.
In addition, 
by allowing the particles to move,
one can model, \eg,
the spatial separation of opinions \cite{schweitzer}.

The Mexican wave (also called La Ola),
is produced by spectators in a stadium, and
it is a well-known example of an instantaneous collective decision.
Since its
direction of motion is spontaneously selected after a rapid
collective decision based on information of limited complexity, it can serve
as a paradigm for similar processes. 
Below, we first give
an overview of general observations and data collected on Mexican
waves in our online survey.
This is followed by a detailed description of
the simulation model we have used including 
its local and global versions
and later the incorporation of
the observed additional left-right asymmetry.
Our results on the spontaneous symmetry breaking transition
and the additional left-right asymmetry
are in Sections \ref{sec:spont} and \ref{sec:left-right}.
Additional calculations and the abbreviated list of data we have used
(from videos and our online survey) are provided in the Appendixes.
Note that the full data set
is available in the preprint version of this paper
from the ArXiv.org server.
The same complete data set together with 
our simulation and evaluation programs can be
downloaded from the supplementary website of this paper at
{\color{blue}{\uline{http://angel.elte.hu/localglobal}}}.

\section{Observations and data on Mexican waves}

\subsection{General observations}

The Mexican wave is launched by a small group of people,
each of them standing up within a short time interval, 
raising their hands high above their heads and then sitting down. 
As this motion is repeated consecutively by groups of close neighbors, 
within a few seconds a stable, linear wave with constant amplitude,
width and speed develops.
Among the reports known to us (see Appendix C) %\cite{survey}
the symmetric solution, \ie, 
two waves started by the same source and
moving in opposite directions, occurs rarely and only by the
coordinated action of an experienced group. 
In a stadium
one source can usually
trigger only one wave moving either left or right, and
the direction 
becomes clear 
short after the initiation.
The mechanism of this rapid self-organizing process
can serve as a paradigm for 
situations 
involving limited interaction
and the selection of one option out of a 
small number of possible choices, \eg,
route choice behaviour in vehicle traffic
\cite{nagel}
or the selection of exits 
during pedestrian escape panic \cite{panic}.

While the Mexican wave rolls,
spectators try to predict when the (nearest) wave will arrive at
their seats and leap to their feet at that moment.
During stationary propagation around the stadium,
both the wave and its velocity are well-defined and spectators can
easily synchronize themselves to the wave's arrival time.
During the short time interval of the initiation, however, the wave's 
direction is not yet known.
Those who can see that
the region of 
active persons is moving towards them will be 
more likely to participate than those who
find that this region is moving away from them. 
In other words, a person is activated by the
combination of two effects:
(i) many of the neighbors are already active and
(ii) the nearest active region (wave) is approaching.
The first is a short range effect, 
while the second is a long range effect.

\subsection{Data on Mexican waves from videos and our online survey}
\label{sec:data}

We have evaluated $15$ recorded waves from videos
and used an online survey to
collect additional information about Mexican waves
(see Appendix C for details).
The waves on the videos were all one-directional,
$7$ of them propagating in the clockwise and $8$ in the
counter-clockwise direction.
From the $75$ visitors of our online survey 
$46$ stated that
the wave's preferred direction was clockwise (right to left), 
$18$ that it was counter-clockwise (left to right), and
$11$ mentioned no preferred direction.

An optional question was the geographical location of
the visitor. Interestingly, 
the ratio of votes for clockwise vs. counter-clockwise was 
$32:12$ from North America
and $5:0$ from Europe, 
while the same ratio was $2:5$ from Australia.
In addition, some visitors explained in detail that the direction
of the wave strongly depends on the relative position of 
initiators compared to each other
and the position of obstacles close to the triggering group
(see, \eg, the answers of visitors No. 29 and 73 
to the question ``Does the wave have a preferred direction?'' 
in Appendix C).
The interaction between the spectators in the stadium was claimed to be 
local, global or both by $15$, $37$ and $17$ people, respectively.
From these answers we concluded that
(i) usually the wave's motion 
is influenced by both 
short range (watching one's neighbors)
and long range interactions (watching the wave as a whole), and
(ii) in most cases an additional left-right 
asymmetry is also present in the system.

\section{Modelling the Mexican wave}

As in Ref.~\cite{briefcomm} we apply an 
{\it excitable medium model} 
to describe the Mexican wave
with each
particle representing 
one spectator in the stadium. 
However, while in Ref.~\cite{briefcomm}
we concentrated on the propagation of the wave,
in the present model we intend to 
capture
the main features of the initiation period.
Our description is inspired by the
Greenberg-Hastings (GH) model of an excitable 
medium \cite{ghmodel}.
In the GH model at the beginning of the simulation 
each particle is in the excitable 
(also called resting or activable) 
state. If at time $t$ there 
is a sufficient number of active particles among the $i$th
particle's neighbors, then the $i$th particle becomes active (excited)
at the next time step, at time $t+\Delta t$. 
After it has been activated, the GH particle deterministically
steps through the $n_{\rm a}$ active states, and then the $n_{\rm r}$
refracter states before it returns to its original, excitable state.
A particle can be activated only if 
it is in the excitable state, and only active particles influence
other particles.

\subsection{Simulation details}

For the simulations, the
stadium 
was folded out to a rectangular lattice with $L_x \times L_y$    
lattice sites (seats). Rows of seats 
became
parallel to the $x$ axis,
boundaries 
were
periodic in the $x$ direction and non-periodic in the $y$
direction. The positive $x$ direction in this coordinate system
corresponds to the clockwise direction in the stadium, while the $y$ axis
is pointing out from the stadium. Stadia for major sport events
usually hold $20,000-80,000$ seats with $60-100$ rows;
in the simulations we have used 
$L_x=400$ and $L_y=80$
(corresponding to $32,000$ spectators)
as a representative size.

In the simulations
the excitable state
corresponds to a person sitting and ready to take part in the wave. Active
states correspond to raising hands when standing up.
Refracter states represent
sitting down plus the time interval when the person is already sitting, 
but not yet activable again. 
In real situations, 
in addition to the active and refracter
states, there is a short, but finite
delay between the time when a person decides to move and the time when
he/she actually starts to move. This delay is due to one's reaction
time, and we model it by inserting $n_{\rm d}$ 
``delay'' states between the
excitable state and the active states. 
In summary, in our model after 
activation a particle is first ``waiting'' 
for $n_{\rm d}$ time steps, 
it is active during the next $n_{\rm a}$ time steps, 
then refracter for $n_{\rm r}$ time steps,
and then it returns to the excitable state.

In one simulation update (one time step)
the $i$th particle is activated with probability $0<p<1$,
\ie, moved from state $0$ (excitable state) 
to state $1$ (delay state),
if (a) it is currently in the excitable state
and (b) the total activation effect, $W_i$, 
on this particle exceeds the activation threshold, $C$. 
The total activation effect, $W_i$, acting on the $i$th particle
is a combination of local (short range) 
and global (long range) interactions:
\bea
W_i = G_i \sum_{j\not= i\atop j\,{\rm active}} w_{j\rightarrow i}\, ,
\label{eq:Wi}
\eea
\noindent
where $G_i$ is the global interaction strength for the $i$th particle
and the sum contains the local effect,
$w_{j\rightarrow i}$,
of each nearby active particle, $j$,
on the $i$th particle.

In the simulations for each particle along the
$y=L_y/2$ line, the time of the particle's first activation, 
was saved as a function of the particle's horizontal
coordinate, $x$.
After triggering a wave
the survival time, $t_S$, of the wave 
was defined as the time below which the 
first activation times showed an increasing function when moving
away from the initiating spot both left and right.

\subsection{Local version of the model}

 If we ignore the global interactions in 
 Eq. (\ref{eq:Wi}), \ie, we set $G_i=1$ for each $i$, then 
\bea
W_i = \sum_{j\not= i\atop j\,{\rm active}} w_{j\rightarrow i} \, .
\label{eq:local}
\eea
 \noindent 
 This is the {\it local version} of the model.
 A simple form for an isotropic,
 exponentially decaying {\it local interaction}
 with characteristic length $R$ is
\bea
w_{j\rightarrow i} = K_i^{-1}\, \e^{ - |\rvec_{ij}|\,/\,R } \, ,
\label{eq:wlocal}
\eea
 \noindent
 where $K_i = \sum_{m} \e^{ - |\rvec_{im}|\,/\,R }$ 
 is a normalizing constant, and
 for any given particle, $i$,
 the summation goes for all $m$ ($m\not= i$) particles.

 In a deterministic case when spectators are identical,
 the excitable $i$th particle is activated, if the sum, $W_i$, 
 of the {\it local} weights exceeds the activation threshold, $C$.
 On the other hand, the activation of spectators in a stadium 
 -- just like most processes involving living systems -- 
 is {\it not entirely deterministic}.
 In the present model 
 this noisy component is taken into account by using a
 stochastic activation rule for each person:
 the activation threshold, $C$, 
 is the same for each particle,
 but the activation of a particle is not deterministic.
 If for the $i$th particle
 the total activation effect, 
 $W_i$, is above the activation threshold, $C$, then
 this particle is activated in the current time step 
 with probability $p$ ($0<p<1$).
 This gives a different response times for each particle.

\subsection{Global version of the model}

Started with a small group of active particles,
the above {\it isotropic local} 
version of the model produces 
-- after a transient circular wave phase --
two symmetric waves propagating in opposite
directions away from the triggering center.
However, all video recordings available to us show
and an overwhelming majority of 
our online visitors report 
that already short after the triggering event 
only one wave is present.
Therefore, an intriguing question is
how one of the two waves is 
suppressed and the 
other is selected so rapidly.
The key effect in selecting the wave's direction 
in the model so quickly is the 
{\it long range interaction:} 
if the active region (perturbation, wave) is moving 
towards (away from) a particle, 
then this will make the activation of that particle more (less) likely.

To take long range interactions into account,
we computed the average $x$ distance of active
particles from the $i$th particle, $x_i^{\,({\rm a})}$,
using an exponentially decaying weight factor:
\bea
x_i^{\,({\rm a})} = \
{ \sum^{\,({\rm a})} \Delta x_{ij} \, \e^{-\Delta x_{ij} / X} \
\over \sum^{\,({\rm a})} \e^{-\Delta x_{ij} / X} } \, .
\label{eq:avgxdact}
\eea
\noindent
Here the 
horizontal distance between 
the $i$th and $j$th particles, $\Delta x_{ij}$, 
%is measured using the periodic boundaries: it 
is the shorter of
the two possible distances allowed by the periodic boundary.
The characteristic length of the long range interaction is $X$
($X\gg R$),
and the summation goes for active $j$ ($j\not= i$) particles.
Denoting by $v_i^{\,({\rm a})}$ the time derivative of $x_i^{\,({\rm a})}$
and by $S$ the sensitivity of spectators to this velocity,
the long range interaction term is
\bea
G_i\, \left(v_i^{\,({\rm a})}\right) =
\cases {
1,& if $v_i^{\,({\rm a})} < 0$;\cr
\e^{-S v_i^{\,({\rm a})}},& if $v_i^{\,({\rm a})} \ge 0$.}
\label{eq:global_term}
\eea
\noindent
Note that $v_i^{\,({\rm a})}$ 
-- the velocity of the active region 
as seen from the $i$th person --
is positive,
if the active region is moving away from this person, and it is
negative, if the active region is approaching the $i$th person.
In the $S\to 0$ limit, $G_i$ is a step function
and the decision about the direction of the wave is very sharp.
In this case one of the two directions is selected quickly and
the wave in the other direction is suppressed, 
because particles on that side 
``measure'' a 
positive
$v_i^{\,({\rm a})}$,
and, consequently, $G_i=0$ and $W_i=0$.
In the $S\to\infty$ limit the global interaction 
term will be a constant,
$G_i=1$, which gives the local version of the model
(Eqs. (\ref{eq:local}) and (\ref{eq:wlocal})).
In the $0<S<\infty$ case an approaching wave,
\ie, $v_i^{\,({\rm a})} < 0$,
will make the $i$th person
more likely to participate, 
while a departing wave less likely.

\subsection{Default parameter values}
\label{sec:default_parameter_values}

Simulation updates were parallel and synchronized, 
and the time step was constant, $0.1s$.
We started each simulation by triggering a small group of
particles: for each particle inside a circle with radius 
$\rho=3$ and centered at $(L_x/2,L_y/2)$, 
we selected a (discrete) time point
randomly from the interval $(0s,1s)$, and moved 
the particle at this time point from state $0$ (excitable state) 
to state $1$ (delay state).
The time derivative, $v_i^{\,({\rm a})}$, 
was computed with Euler's formula 
and a time step of $\Delta t=0.5s$. 
If at time $t$ for the $i$th particle either 
$x_i^{\,({\rm a})}(t)$ or $x_i^{\,({\rm a})}(t-\Delta t)$ was not available,
then the long range interaction coefficient, 
$G_i\left(v_i^{\,({\rm a})}\right)$,
was replaced by $1$.
Default parameter values were
$R=3$, $X=100$, $C=0.23$, $p=0.9$, 
$n_{\rm d}=1$, $n_{\rm a}=10$, $n_{\rm r}=20$.
Distances were measured in seats, 
time was measured in seconds.
To compute the local weights,
 we applied a cutoff and restricted 
 the summation to $|\rvec_{ij}| \le 3R$.
%Note that because of the translational symmetry of the system 
% in the $x$ (``horizontal'') direction $K_i$ depends 
% only on the $y$ (``vertical'') coordinate of a particle.
%Data points of the figures show averages over $1,000$ simulations.

\section{Spontaneous symmetry breaking}
\label{sec:spont}

\begin{figure}[t!]
\centerline{\resizebox{0.9\textwidth}{!}{\rotatebox{0}{\includegraphics{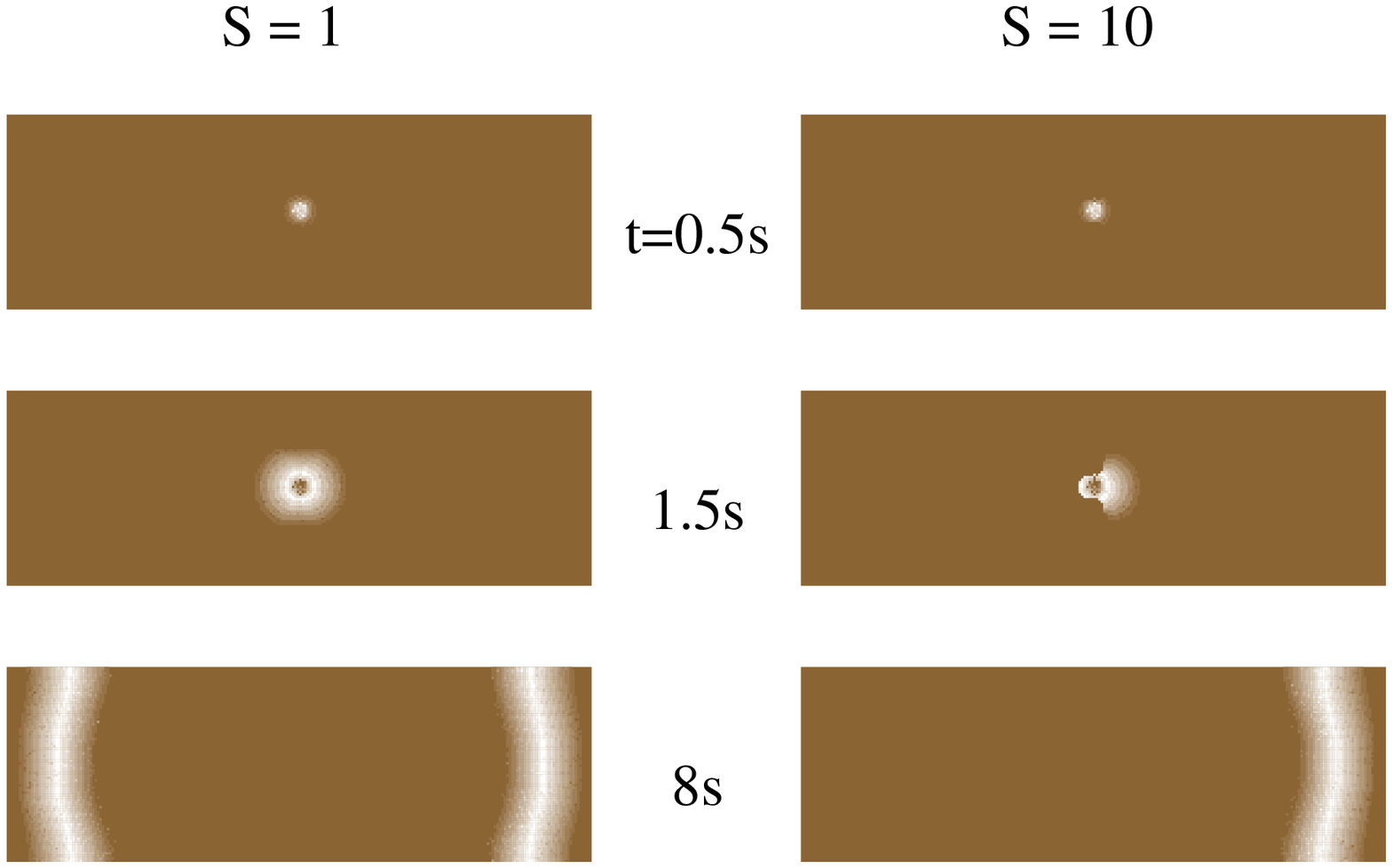}}}}
%\centerline{\includegraphics[angle=0,width=0.45\columnwidth]{FIG1_waveStates.ps}}
%\centerline{\includegraphics[angle=0,width=0.9]{FIG1_waveStates.ps}}
%%% figure caption word count: 141
\caption{Spontaneous 
symmetry breaking in the Mexican wave simulations:
each spectator is represented as one particle 
in a rectangular lattice.
Shown are parts of the simulation area at $0.5s$, $1.5s$ and $8s$
after the triggering event 
(see Section \ref{sec:default_parameter_values} 
for the default parameter values).
Excitable particles are colored dark.
The increasing color brightness of active particles represents the
different stages while standing up, 
the decreasing color brightness of the
first $10$ refracter states represents the stages of sitting down, 
while the dark color of the remaining $10$ refracter states 
indicates that the person is already
sitting, though not yet activable again.
{\bf Left column.} 
If the control parameter, $S$
(which is also the relative weight of global interactions),
is low, then the stable solution contains two waves moving 
in opposite directions. 
{\bf Right column.} 
At higher values of the control parameter
the asymmetric
solution -- one wave moving either left or right --
becomes stable.
}
\label{fig:waveStates}
\end{figure}

In our model the relative 
weight of global interactions is given by $S$.
In the $S\to 0$ limit one obtains the isotropic local 
version of the model,
where the stable solution contains two oppositely moving waves.
Raising $S$ causes a spontaneous symmetry breaking 
(global interactions are ``turned on''), 
and the symmetrical solution becomes unstable.
If $S$ is high, then soon after the initiation
one of the two directions is selected and
propagation in the other direction is stopped.
Thus, $S$ changes the symmetry properties of the stable solution
and can be used as a {\it control parameter}.
Figure \ref{fig:waveStates} shows how
at different values of $S$
either two oppositely moving waves
develop from one initiating source or 
one of the two directions is selected.

\begin{figure}[t!]
\centerline{\resizebox{1.2\textwidth}{!}{\rotatebox{-90}{\includegraphics{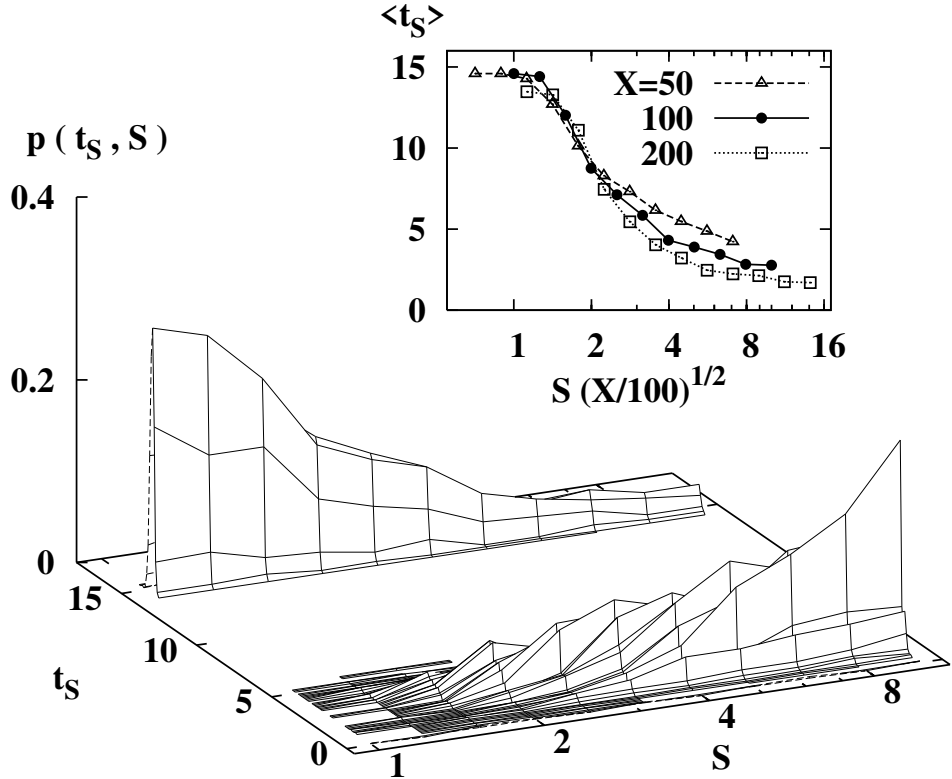}}}}
%\centerline{\includegraphics[angle=-90,width=0.6\columnwidth]{FIG2_dist.ps}}
%\centerline{\includegraphics[angle=-90,width=1.2]{FIG2_dist.ps}}
%%% figure caption word count: 117
\caption{
{\bf Main panel.}
Transition between the symmetric 
(two waves moving in opposite directions) 
and asymmetric solutions (one wave moving either left or right)
in the Mexican wave model.
The control parameter, $S$,
which is the parameter of global asymmetry in the model,
is analogous to the inverse temperature, $\beta$, of
temperature-controlled transitions.
For each $S$ the distribution of $t_S$ 
(``survival time'') is displayed:
$t_S$ is the time until which the two oppositely 
moving waves are both present in the system.
{\bf Inset.}
The average survival time, 
$\langle t_S\rangle$, of the symmetrical solution
for different values of the global interaction length, $X$.
The transition point, $S_C$, scales as $X^{-1/2}$, and the
transition itself becomes sharper with increasing $X$.
Data points show averages over $1,000$ simulations 
for each value of $S$.
}
\label{fig:S}
\end{figure}

Since the control parameter, $S$,
tunes the stability of the symmetric solution,
we have selected an {\it order parameter} measuring this stability.
For a wave starting off asymmetrically at the triggering spot
the survival time is $t_S\approx 0$.
On the other hand, for an infinitely stable symmetrical solution
the survival time is a constant finite value corresponding to the
time needed for the two waves to meet at the 
opposite end of the stadium.
The main panel of Fig.~\ref{fig:S} shows the 
distribution of $t_S$ values as a function of 
the control parameter, $S$. 
We found that for a range of $S$ values
the distribution of $t_S$ has two distinct peaks, which is
analogous
to the distribution of the order parameter 
in the vicinity of the transition point in systems undergoing a 
discontinuous transition.
The two phases are
the symmetric solution with two waves (high $t_S$) 
and the asymmetric solution,
a wave moving either left or right (low $t_S$).
Since the size of the initiating group is always 
finite (in the range of a few dozen particles), 
the phenomenon is inherently mesoscopic and 
the transition is not very sharp.

The inset of Fig.~\ref{fig:S} shows the average of the order
parameter, $\langle t_S\rangle$,
as a function of $S$
at different values of the long range 
interaction length, $X$.
The transition point
scales as $X^{-1/2}$, and the transition becomes sharper with
increasing $X$.
This 
scaling is caused by (i) the 
uncorrelated random activation of particles
-- both in space and time --
in the triggering group
and (ii) the locally linear shape
of the global interaction term 
as a function of the horizontal coordinate 
(see Eq. (\ref{eq:global_term})).
A detailed derivation of this scaling is provided in Appendix B.
During the triggering the expected speed 
of the active region, $v_i^{\,({\rm a})}$,
-- as seen from a nearby excitable particle, $i$ --
scales as $X^{1/2}$. Since the global 
interaction term is a function of $S v_i^{\,({\rm a})}$,
this gives $S_C\sim X^{-1/2}$
(see Appendix B for details).

\section{Left-right symmetry breaking}
\label{sec:left-right}

The above description assumes that during the triggering of the wave,
the participating spectators stand up in an uncorrelated fashion.
In real situations, however, additional, very fast interactions are
present, and the decision about the direction of the wave is usually
faster than predicted here.
It is well-known that people inside the triggering 
spot influence each other both before they start moving 
and also during
the very short time interval of the triggering.
From the responses to our online survey (see Appendix C) %\cite{survey}
we found that one important additional effect
influencing the wave's direction and stability
is the local geometry, \eg,
the presence of obstacles around the triggering spot.
Also, people tend to react asymmetrically to disturbances
occurring on their left and right sides;
this is caused by our
physiological asymmetry and our expectations.
If by a combination of these  
effects people react stronger to stimuli on, \eg, 
their left, than to those on their right,
then a wave will most likely propagate 
from the left to the right,
which is the counter-clockwise direction in the 
stadium when watched from above.

To model the inherent local asymmetry during the triggering,
we extended the global version of the model 
and changed Eq. (\ref{eq:wlocal}).
In the local coordinate system of the $i$th particle
let $\varphi$ denote the angle of 
$\rvec_{ij}$ pointing from the 
$i$th particle to the $j$th particle.
If $\rvec_{ij}$ points to the left 
(the clockwise direction when watched from above), then $\varphi=0$, 
and for $\rvec_{ij}$ pointing radially out from the stadium,
$\varphi=\pi/2$.
We used the following direction-dependent local 
interaction term
(compare to Eq. (\ref{eq:wlocal})):
\bea
w_{i\rightarrow j} = \
K_i^{-1}\, \e^{- |\rvec_{ij}| / R}\,
\big[ (1-\delta) + \delta\cos\,(\pi-\varphi) \big] \, .
\label{eq:wlocal2}
\eea
\noindent
Similarly to Eq. (\ref{eq:wlocal}),
$K_i$ is a normalizing constant chosen such 
that for the $i$th particle the sum of 
$w_{i\rightarrow j}$ ($j\not= i$) values is $1$.

\begin{figure}[t!]
\centerline{\resizebox{\textwidth}{!}{\rotatebox{-90}{\includegraphics{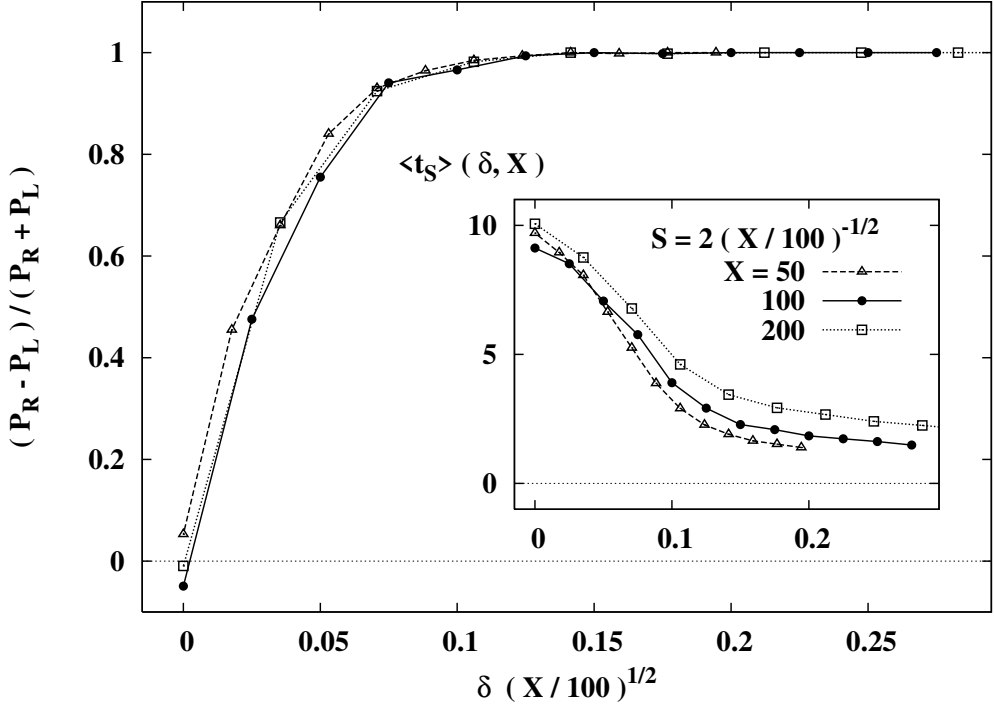}}}}
%\centerline{\includegraphics[angle=-90,width=0.5\columnwidth]{FIG3_delta.ps}}
%\centerline{\includegraphics[angle=-90,width=1\columnwidth]{fig/delta.ps}}
%%% figure caption word count: 105
\caption{Left-right 
symmetry breaking in the Mexican wave simulations
caused by the 
local asymmetry
during the triggering of the wave.
The $\delta=0$ case corresponds to the isotropic local version 
of the 
model (see Eqs. (\ref{eq:local}) and (\ref{eq:wlocal})).
The sensitivity of particles to the speed of the active region is
$S=2\,(X/100)^{-1/2}$, \ie, the system is at the transition point
(see the inset of Fig.~\ref{fig:S}).
{\bf Main panel.}
The difference between the probabilities of 
left 
($P_L$)
and right 
($P_R$)
moving 
waves
as a function of the local left-right 
asymmetry parameter, $\delta$.
{\bf Inset.}
The average survival time,
$\langle t_S\rangle$, 
for different values of the long range interaction 
length, $X$.
Data points show averages over $1,000$ simulations 
for each value of $\delta$.}
\label{fig:delta}
\end{figure}

In the model, ``switching on'' the
left-right asymmetry of local interactions 
has two simultaneous effects.
(i) The balance between left and right moving waves will be
broken (see the main panel of Fig.~\ref{fig:delta}):  
by increasing $\delta$ the probability of the right moving wave,
$P_R$, will be dominant over the probability of the left moving wave,
$P_L$. 
(ii) The spontaneous symmetry breaking transition will be shifted 
(inset of Fig.~\ref{fig:delta}). The inset of
Fig.~\ref{fig:waveStates} shows that for all values of $X$, the
transition point of the spontaneous symmetry breaking is at
approximately $S_C(X)=2(X/100)^{-1/2}$. 
This is the value of $S$, where the
average survival time of the symmetric solution (two waves), 
$\langle t_S \rangle$ reaches half of its maximum value. 
However, with the same 
$S$ values and $\delta>0$, the survival time of the symmetric
solution becomes significantly smaller, \ie, 
the symmetric solution is
destabilized by the left-right asymmetry term.

Observe that in both cases 
the transition curves collapse when plotted 
as a function of $\delta X^{1/2}$, \ie,
the transition point, $\delta_C$,
scales with the global
interaction length as $X^{-1/2}$.
This is the same scaling as the one that we have seen earlier 
for $S_C$ in the spontaneous symmetry breaking, 
and it has the same two reasons 
(see Appendix B for a detailed derivation):
the random activation of particles in the triggering spot is
uncorrelated and the left-right asymmetry term is 
a locally linear function.

\section*{Conclusions}

We have presented a simple realistic model of an 
instantaneous collective human decision process,
where the interplay of local and global interactions
leads to a spontaneous symmetry breaking.
The decision we have modelled 
concerns the direction of propagation 
of the Mexican wave (La Ola) in a stadium 
after a small group of people stands up to initiate the wave.
Although the situation and the model we have considered 
is relatively
simple, they give an insight into the mechanisms by which quick
decisions are made by groups of people. Understanding such phenomena
is important in various contexts including the spreading of
excitement in panicking crowds or during 
collective actions of humans
at gatherings and
during collective financial decisions.

\section*{Acknowledgements}

We thank D. Helbing for permission to use video material and the
visitors of our online questionnaire for their answers.
This work was supported in part by the Hungarian Sci. Res.
Fund (OTKA) Grants No. T049674 and D048422.

%:::::::::::::::::::::::::::::::::::::::::::::::::::::::::::::::::::::::
% References:
%:::::::::::::::::::::::::::::::::::::::::::::::::::::::::::::::::::::::

\vfill\eject

\section*{Appendix A. Simulation software}

The simulation program was run on Linux computers. Its core performing
the numerical operations was written in C++ and the graphical
interface in Qt. The simulation program can be run both with and
without visualization. The additional scripts and utilities that can
start the simulation program at various parameter values and evaluate
the results were written in Perl and C.  
Our programs
-- including their source codes and a short documentation --
can be downloaded from our website at
{\color{blue}{\uline{http://angel.elte.hu/localglobal}}}.

\section*{Appendix B. Scaling of the transition point in the
  spontaneous symmetry breaking} 
To interpret the 
$S_C\sim X^{-1/2}$ behaviour 
(see the inset of Fig.~\ref{fig:S})
consider the $i$th particle,
just outside the initial activation
spot (a circle with radius $\rho$).
The horizontal ($x$) distance of this particle from the 
center of the spot is $\ell$.
When the simulation is started, the $i$th particle 
is affected by the active particles, $j$, inside the spot. 
Since $|\Delta x_{ij}-\ell|\le\rho\ll X$,
the exponential weight function in 
Eq. (\ref{eq:avgxdact}) 
can be approximated with a linear function:
\bea
x_i^{\,({\rm a})} &=& \
{ \
\sum^{\,({\rm a})}\Delta x_{ij}\,\e^{-\ell/X}\,\e^{-(\Delta x_{ij}-\ell) / X} \
\over\sum^{\,({\rm a})} \e^{-\ell/X}\,\e^{-(\Delta x_{ij}-\ell) / X} \
} \
\simeq \
\nonumber \\
&\simeq& \
{ \
\sum^{\,({\rm a})} \Delta x_{ij}\,[1-(\Delta x_{ij}-\ell) / X] \
\over \sum^{\,({\rm a})} [1-(\Delta x_{ij}-\ell) / X] \
} \, .
\label{eq:lin}
\eea
%\bea
%x_i^{\,({\rm a})} = \
%{ \
%\sum^{\,({\rm a})}\Delta x_{ij}\,\e^{-\ell/X}\,\e^{-(\Delta x_{ij}-\ell) / X} \
%\over\sum^{\,({\rm a})} \e^{-\ell/X}\,\e^{-(\Delta x_{ij}-\ell) / X} \
%} \
%\simeq \
%{ \
%\sum^{\,({\rm a})} \Delta x_{ij}\,[1-(\Delta x_{ij}-\ell) / X] \
%\over \sum^{\,({\rm a})} [1-(\Delta x_{ij}-\ell) / X] \
%} \, .
%\label{eq:lin}
%\eea
\noindent
The $\sum^{\,({\rm a})}$ summations run for
active particles, $j$
($j\not= i$), inside the triggering spot.
Denoting $(\Delta x_{ij}-\ell) / X$ by $\eta_{ij}$,
and the average by an overline, one can obtain
\bea
x_i^{\,({\rm a})}-\ell = X - X {\sum^{\,({\rm a})} (1-\eta_{ij})^2 \over \
\sum^{\,({\rm a})} (1-\eta_{ij})} = \
X - X {1-2{\overline\eta}+\overline{\eta^2}\over \
1-{\overline\eta}} \nonumber \, .
\eea
\noindent 
Note that from this point on 
each average containing 
$\eta$ has an additional index, $i$. % not shown in the formulas.
Since $|\eta_{ij}|\ll 1$,
we can drop the
${\mathcal O}\big({\overline\eta}^2\big)$ and
${\mathcal O}\big({\overline{\eta^2}}\big)$ terms:
\bea
x_i^{\,({\rm a})}-\ell \cong \overline\eta\, X \, .
\label{eq:etaX}
\eea
\noindent
During the initial triggering, activations occur
with uniform spatial distribution inside the spot,
and also with uniform distribution in time.
Moreover,
the expected distribution of 
$\eta_{ij}$ values is symmetrical around $0$ at any time point.
Thus, the process of the summation of $\eta_{ij}$ 
values is always a random walk around $0$,
and the expected sum is proportional to the square root of the 
linear scale of the summed values.
The variable part of $\eta_{ij}$ is 
$\Delta x_{ij}/X$, therefore,
the linear scale is proportional to $X^{-1}$, 
and the changing part of the above sum
scales as $X^{-1/2}$. 
Similarly, the changing part of the average, 
$\overline\eta$,
and the expected rate of change of the average,
${d\over dt}\overline\eta$,
are both proportional to $X^{-1/2}$.
Differentiating Eq. (\ref{eq:etaX}) with respect to time gives
$v_i^{\,({\rm a})}$ on the left and 
$X\,{d\over dt}\overline\eta$ on the 
right hand side, 
and we get 
$v_i^{\,({\rm a})}\sim X^{1/2}$.
The global 
interaction term is a function of $S v_i^{\,({\rm a})}$,
and so the product $S_C\,v_i^{\,({\rm a})}$
should be constant when 
$X$ is changed, therefore, $S_C\sim X^{-1/2}$.

The above approximation for the speed of the active region,
$v_i^{\,({\rm a})}$, is valid only during the triggering.
The $v_i^{\,({\rm a})}\sim X^{1/2}$ proportionality means
that 
(i) the decision process is slightly faster for a larger global
interaction length and 
(ii) with a weak global interaction ($X\to 0$)
there is no spontaneous symmetry breaking.

\section*{Appendix C. Data on Mexican waves}

\subsection*{Evaluation of recorded Mexican waves (videos)} 

\medskip
\begin{table}[!t]
  \begin{center}
    \begin{tabular}{|l|l|l|l|}\hline
      \rowcolor[rgb]{0.8,0.8,0.8}
      {\bf {\small File name}} & {\bf Duration [s]} & {\bf Direction} & {\bf Source}\\ \hline
laola01.mpeg & 14 & counter-clockwise & D.H. \\ \hline
laola02.mpeg & 7  & clockwise         & D.H. \\ \hline
laola03.mpeg & 10 & clockwise         & D.H. \\ \hline
laola04.mpeg & 11 & clockwise         & D.H. \\ \hline
laola05.mpeg & 10 & clockwise         & D.H. \\ \hline
laola06.mpeg & 8  & counter-clockwise & D.H. \\ \hline
laola13.mpeg & 7  & clockwise         & D.H. \\ \hline
laola14.mpeg & 14 & clockwise         & D.H. \\ \hline
laola15.mpeg & 14 & counter-clockwise & D.H. \\ \hline
laola16.mpeg & 14 & counter-clockwise & D.H. \\ \hline
laola17.mpeg & 14 & counter-clockwise & D.H. \\ \hline
laola21.mpeg & 2  & counter-clockwise & D.H. \\ \hline
laola22.mpeg & 6  & counter-clockwise & D.H. \\ \hline
Mvc-528w.mpg & 12 & clockwise         & web1 \\ \hline
sportwav.MPG & 5  & counter-clockwise & web2 \\ \hline
    \end{tabular}
\medskip
\caption{Direction and duration of $15$
recorded Mexican waves evaluated from videos. 
See Section \ref{sec:data} of the paper and Appendix C
for a detailed analysis.
Sources: D.H.: Dirk Helbing, web1: derat.nl, web2:
web.ukonline.co.uk/Members/s.livingston.}
  \end{center}
%\label{table:videos}
\end{table}
\medskip

Table 1 lists the $15$ recorded 
Mexican waves we have evaluated:
$14$ of them were waves rolling among the spectators
in stadia and indoor halls built for athletic events and holding 
up to $50,000$ spectators, 
while $1$ of the waves 
(the file is called sportwav.MPG) was performed
on a sports field by a group of approximately $50$ children.
On all videos only one wave -- moving in one of the two possible
directions -- can be seen, in 
$7$ cases it was a clockwise wave and in the remaining $8$ cases 
it was moving in the counter-clockwise direction.

\subsection*{Answers collected in our online survey}

We have set up an online survey on Mexican waves
at {\color{blue}{\uline{http://angel.elte.hu/wave}}}.
$\rightarrow$ select {\it SURVEY} in the top right corner.
The answers to this survey between July 2004 and August 2005 
are listed below.
The question about the geographical location of the visitor was added
in September, 2004.
A detailed evaluation of the answers is given 
in Section \ref{sec:data} above.

\medskip
{\tiny
\begin{itemize}

\subsubsection*{Answers 1 to 10 in the online survey}
\smallskip
\item[] {\bf \uline{(1) Jul 14, 2004}}\begin{itemize}
\item[*]{\bf Have you ever taken part in a Mexican wave?}
\\
 YES 
\item[*]{\bf If yes, did you follow your neighbours or you considered rather the motion of the wave as a whole? Please, explain below.}
\\
As a whole. 
\item[*]{\bf Does the wave have a preferred direction? (Clockwise or counter-clockwise?)}
\\
yes. clockwise. 
\item[*]{\bf Have you ever seen two waves initiated by a single source? (i.e., the wave leaving in both directions) If yes, please give details, e.g., when and where.}
\\
dont remember. 
\item[*]{\bf Have you ever taken part in initiating a wave? What did you do? How many of you were needed? Did you try to influence the direction of the wave from the very beginning?}
\\
no. 
\end{itemize}
\smallskip
\item[] {\bf \uline{(2) Jul 18, 2004}}\begin{itemize}
\item[*]{\bf Have you ever taken part in a Mexican wave?}
\\
 YES 
\item[*]{\bf If yes, did you follow your neighbours or you considered rather the motion of the wave as a whole? Please, explain below.}
\\
I would tend to anticipate the movement, but would move at the same time as my immediate neighbors. Im the wave-zealous type who wants to keep it going as long as possible. 
\item[*]{\bf Does the wave have a preferred direction? (Clockwise or counter-clockwise?)}
\\
Most of the ones Ive seen are clockwise. 
\item[*]{\bf Have you ever seen two waves initiated by a single source? (i.e., the wave leaving in both directions) If yes, please give details, e.g., when and where.}
\\
Never seems to be stable; one side gives up quickly. 
\item[*]{\bf Have you ever taken part in initiating a wave? What did you do? How many of you were needed? Did you try to influence the direction of the wave from the very beginning?}
\\
Yes. Ive been in the small group trying to start one. It was not the size of the group, but the density that seemed to matter. 
\end{itemize}
\smallskip
\item[] {\bf \uline{(3) Aug 1, 2004}}\begin{itemize}
\item[*]{\bf Have you ever taken part in a Mexican wave?}
\\
 YES 
\item[*]{\bf If yes, did you follow your neighbours or you considered rather the motion of the wave as a whole? Please, explain below.}
\\
you can see it coming to your section and it is easy to see when to start and stop. 
\item[*]{\bf Does the wave have a preferred direction? (Clockwise or counter-clockwise?)}
\\
clockwise, not sure why, but I dont beleive Ive ever seen it go counter clockwise 
\item[*]{\bf Have you ever seen two waves initiated by a single source? (i.e., the wave leaving in both directions) If yes, please give details, e.g., when and where.}
\\
no 
\item[*]{\bf Have you ever taken part in initiating a wave? What did you do? How many of you were needed? Did you try to influence the direction of the wave from the very beginning?}
\\
no 
\end{itemize}
\smallskip
\item[] {\bf \uline{(4) Aug 24, 2004}}\begin{itemize}
\item[*]{\bf Have you ever taken part in a Mexican wave?}
\\
 YES 
\item[*]{\bf If yes, did you follow your neighbours or you considered rather the motion of the wave as a whole? Please, explain below.}
\\
I followed the wave as a whole 
\item[*]{\bf Does the wave have a preferred direction? (Clockwise or counter-clockwise?)}
\\
Clockwise 
\item[*]{\bf Have you ever seen two waves initiated by a single source? (i.e., the wave leaving in both directions) If yes, please give details, e.g., when and where.}
\\
no 
\item[*]{\bf Have you ever taken part in initiating a wave? What did you do? How many of you were needed? Did you try to influence the direction of the wave from the very beginning?}
\\
no 
\end{itemize}
\smallskip
\item[] {\bf \uline{(5) Aug 24, 2004}}\begin{itemize}
\item[*]{\bf Have you ever taken part in a Mexican wave?}
\\
 YES 
\item[*]{\bf If yes, did you follow your neighbours or you considered rather the motion of the wave as a whole? Please, explain below.}
\\
wave as a whole 
\item[*]{\bf Does the wave have a preferred direction? (Clockwise or counter-clockwise?)}
\\
clockwise 
\item[*]{\bf Have you ever seen two waves initiated by a single source? (i.e., the wave leaving in both directions) If yes, please give details, e.g., when and where.}
\\
--
\item[*]{\bf Have you ever taken part in initiating a wave? What did you do? How many of you were needed? Did you try to influence the direction of the wave from the very beginning?}
\\
--
\end{itemize}
\smallskip
\item[] {\bf \uline{(6) Aug 25, 2004}}\begin{itemize}
\item[*]{\bf Have you ever taken part in a Mexican wave?}
\\
 YES 
\item[*]{\bf If yes, did you follow your neighbours or you considered rather the motion of the wave as a whole? Please, explain below.}
\\
followed along with the wave 
\item[*]{\bf Does the wave have a preferred direction? (Clockwise or counter-clockwise?)}
\\
clockwise 
\item[*]{\bf Have you ever seen two waves initiated by a single source? (i.e., the wave leaving in both directions) If yes, please give details, e.g., when and where.}
\\
never 
\item[*]{\bf Have you ever taken part in initiating a wave? What did you do? How many of you were needed? Did you try to influence the direction of the wave from the very beginning?}
\\
about a dozen people started a wave with one or two stranding in front facing us. they counted 1....2....3...then we started it. it just always goes clockwise 
\end{itemize}
\smallskip
\item[] {\bf \uline{(7) Sep 28, 2004}}\begin{itemize}
\item[*]{\bf Have you ever taken part in a Mexican wave?}
\\
 YES 
\item[*]{\bf If yes, did you follow your neighbours or you considered rather the motion of the wave as a whole? Please, explain below.}
\\
Followed not the neighbor but the neighbors about 1/8 wavelength away 
\item[*]{\bf Does the wave have a preferred direction? (Clockwise or counter-clockwise?)}
\\
Yes. Moving from the people standing to the people sitting 
\item[*]{\bf Have you ever seen two waves initiated by a single source? (i.e., the wave leaving in both directions) If yes, please give details, e.g., when and where.}
\\
No 
\item[*]{\bf Have you ever taken part in initiating a wave? What did you do? How many of you were needed? Did you try to influence the direction of the wave from the very beginning?}
\\
Yes. Successful rarely. Tried to encourage people in a given direction to move. 
\item[*]{\bf Please, select your geographical location.}
\\
North America 
\end{itemize}
\smallskip
\item[] {\bf \uline{(8) Sep 29, 2004}}\begin{itemize}
\item[*]{\bf Have you ever taken part in a Mexican wave?}
\\
 YES 
\item[*]{\bf If yes, did you follow your neighbours or you considered rather the motion of the wave as a whole? Please, explain below.}
\\
Considewred the motion as a whole even to the extent of being out of synch with the neighbours when they have got out of time 
\item[*]{\bf Does the wave have a preferred direction? (Clockwise or counter-clockwise?)}
\\
Yes, Clockwise 
\item[*]{\bf Have you ever seen two waves initiated by a single source? (i.e., the wave leaving in both directions) If yes, please give details, e.g., when and where.}
\\
No 
\item[*]{\bf Have you ever taken part in initiating a wave? What did you do? How many of you were needed? Did you try to influence the direction of the wave from the very beginning?}
\\
Yes, about 6-10 people Not conciously, but we tended to look to our right to start them up 
\item[*]{\bf Please, select your geographical location.}
\\
Australia 
\end{itemize}
\smallskip
\item[] {\bf \uline{(9) Sep 29, 2004}}\begin{itemize}
\item[*]{\bf Have you ever taken part in a Mexican wave?}
\\
 YES 
\item[*]{\bf If yes, did you follow your neighbours or you considered rather the motion of the wave as a whole? Please, explain below.}
\\
I considered the motion of the wave as a whole. 
\item[*]{\bf Does the wave have a preferred direction? (Clockwise or counter-clockwise?)}
\\
Clockwise 
\item[*]{\bf Have you ever seen two waves initiated by a single source? (i.e., the wave leaving in both directions) If yes, please give details, e.g., when and where.}
\\
No. 
\item[*]{\bf Have you ever taken part in initiating a wave? What did you do? How many of you were needed? Did you try to influence the direction of the wave from the very beginning?}
\\
No. 
\item[*]{\bf Please, select your geographical location.}
\\
Africa 
\end{itemize}
\smallskip
\item[] {\bf \uline{(10) Sep 29, 2004}}\begin{itemize}
\item[*]{\bf Have you ever taken part in a Mexican wave?}
\\
 YES 
\item[*]{\bf If yes, did you follow your neighbours or you considered rather the motion of the wave as a whole? Please, explain below.}
\\
Wtached the wave. Was aware of its progress around the ground. Was anticipating and preparing for the crest. (I guess that proves Im not a cellular automaton.) 
\item[*]{\bf Does the wave have a preferred direction? (Clockwise or counter-clockwise?)}
\\
No. 
\item[*]{\bf Have you ever seen two waves initiated by a single source? (i.e., the wave leaving in both directions) If yes, please give details, e.g., when and where.}
\\
No. 
\item[*]{\bf Have you ever taken part in initiating a wave? What did you do? How many of you were needed? Did you try to influence the direction of the wave from the very beginning?}
\\
No 
\item[*]{\bf Please, select your geographical location.}
\\
Australia 
\end{itemize}

\subsubsection*{Answers 11 to 20 in the online survey}
\medskip
\smallskip
\item[] {\bf \uline{(11) Sep 29, 2004}}\begin{itemize}
\item[*]{\bf Have you ever taken part in a Mexican wave?}
\\
 YES 
\item[*]{\bf If yes, did you follow your neighbours or you considered rather the motion of the wave as a whole? Please, explain below.}
\\
motion of the wave 
\item[*]{\bf Does the wave have a preferred direction? (Clockwise or counter-clockwise?)}
\\
clockwise 
\item[*]{\bf Have you ever seen two waves initiated by a single source? (i.e., the wave leaving in both directions) If yes, please give details, e.g., when and where.}
\\
yes - chennai india 
\item[*]{\bf Have you ever taken part in initiating a wave? What did you do? How many of you were needed? Did you try to influence the direction of the wave from the very beginning?}
\\
no 
\item[*]{\bf Please, select your geographical location.}
\\
North America 
\end{itemize}
\smallskip
\item[] {\bf \uline{(12) Sep 30, 2004}}\begin{itemize}
\item[*]{\bf Have you ever taken part in a Mexican wave?}
\\
 YES 
\item[*]{\bf If yes, did you follow your neighbours or you considered rather the motion of the wave as a whole? Please, explain below.}
\\
Motion of the wave .. since the ground was an oval i could see the propagation of wave from my sides and i timed my up to match the wave. But it didnt go too well.. 
\item[*]{\bf Does the wave have a preferred direction? (Clockwise or counter-clockwise?)}
\\
No. But to me left to right seems good. 
\item[*]{\bf Have you ever seen two waves initiated by a single source? (i.e., the wave leaving in both directions) If yes, please give details, e.g., when and where.}
\\
No. 
\item[*]{\bf Have you ever taken part in initiating a wave? What did you do? How many of you were needed? Did you try to influence the direction of the wave from the very beginning?}
\\
No. 
\item[*]{\bf Please, select your geographical location.}
\\
North America 
\end{itemize}
\smallskip
\item[] {\bf \uline{(13) Sep 30, 2004}}\begin{itemize}
\item[*]{\bf Have you ever taken part in a Mexican wave?}
\\
 YES 
\item[*]{\bf If yes, did you follow your neighbours or you considered rather the motion of the wave as a whole? Please, explain below.}
\\
Both. Mainly neighbors, but if they were not well-timed, then I put more emphasis on the wave as a whole. 
\item[*]{\bf Does the wave have a preferred direction? (Clockwise or counter-clockwise?)}
\\
Almost always clockwise in the USA. Cheerleaders or yell leaders can often get the crowd organized to do different things however: different directions, multiple waves at once, different waves on different levels of the statium, etc. 
\item[*]{\bf Have you ever seen two waves initiated by a single source? (i.e., the wave leaving in both directions) If yes, please give details, e.g., when and where.}
\\
No, although Ive read about people doing it. I think it would take a cheerleader, as above. 
\item[*]{\bf Have you ever taken part in initiating a wave? What did you do? How many of you were needed? Did you try to influence the direction of the wave from the very beginning?}
\\
No. I do not like the wave and do not participate in it anymore, as it is distracting and irrelevant to the sporting event. I encourage you to use a pair of names for the term; although people outside of North American may call it the Mexican wave due to the 1986 World Cup exposure, Americans had been doing it, and calling it simply the Wave,for years prior to that. That is, this stadium activity had been often performed, and called the Wave, for years before 1986. Granted, Americans screw things up by calling football soccer. But analogously, Europeans are screwing things up by calling the Wave the Mexican wave. 
\item[*]{\bf Please, select your geographical location.}
\\
North America 
\end{itemize}
\smallskip
\item[] {\bf \uline{(14) Sep 30, 2004}}\begin{itemize}
\item[*]{\bf Have you ever taken part in a Mexican wave?}
\\
 YES 
\item[*]{\bf If yes, did you follow your neighbours or you considered rather the motion of the wave as a whole? Please, explain below.}
\\
watched it all around the stadium. You must consider the wave as a holistic event 
\item[*]{\bf Does the wave have a preferred direction? (Clockwise or counter-clockwise?)}
\\
clockwise 
\item[*]{\bf Have you ever seen two waves initiated by a single source? (i.e., the wave leaving in both directions) If yes, please give details, e.g., when and where.}
\\
no 
\item[*]{\bf Have you ever taken part in initiating a wave? What did you do? How many of you were needed? Did you try to influence the direction of the wave from the very beginning?}
\\
no 
\item[*]{\bf Please, select your geographical location.}
\\
Australia 
\end{itemize}
\smallskip
\item[] {\bf \uline{(15) Sep 30, 2004}}\begin{itemize}
\item[*]{\bf Have you ever taken part in a Mexican wave?}
\\
 YES 
\item[*]{\bf If yes, did you follow your neighbours or you considered rather the motion of the wave as a whole? Please, explain below.}
\\
Followed the neighbours 
\item[*]{\bf Does the wave have a preferred direction? (Clockwise or counter-clockwise?)}
\\
Wherever I have participated it has been anticlockwise. That is, wave comes from my left and proceeds to right. 
\item[*]{\bf Have you ever seen two waves initiated by a single source? (i.e., the wave leaving in both directions) If yes, please give details, e.g., when and where.}
\\
No 
\item[*]{\bf Have you ever taken part in initiating a wave? What did you do? How many of you were needed? Did you try to influence the direction of the wave from the very beginning?}
\\
No 
\item[*]{\bf Please, select your geographical location.}
\\
Asia 
\end{itemize}
\smallskip
\item[] {\bf \uline{(16) Oct 8, 2004}}\begin{itemize}
\item[*]{\bf Have you ever taken part in a Mexican wave?}
\\
 YES 
\item[*]{\bf If yes, did you follow your neighbours or you considered rather the motion of the wave as a whole? Please, explain below.}
\\
Wave as a whole. ie I waited until the majority of people in the section beofre me had risen and sat and then I got up and rose, chudked all sorts of shit in the air and then sat. 
\item[*]{\bf Does the wave have a preferred direction? (Clockwise or counter-clockwise?)}
\\
anti clockwise 
\item[*]{\bf Have you ever seen two waves initiated by a single source? (i.e., the wave leaving in both directions) If yes, please give details, e.g., when and where.}
\\
no 
\item[*]{\bf Have you ever taken part in initiating a wave? What did you do? How many of you were needed? Did you try to influence the direction of the wave from the very beginning?}
\\
Yes, basically a group of ten or more will deicde to initiate. Then they will count down from ten. Then the bloke on the left or giht will rise depending upon the direction they want it to go , and then the bloke next to him will rise, followed by the guy next to him. Then everyone follows. 
\item[*]{\bf Please, select your geographical location.}
\\
Australia 
\end{itemize}
\smallskip
\item[] {\bf \uline{(17) Oct 10, 2004}}\begin{itemize}
\item[*]{\bf Have you ever taken part in a Mexican wave?}
\\
 YES 
\item[*]{\bf If yes, did you follow your neighbours or you considered rather the motion of the wave as a whole? Please, explain below.}
\\
Just to let it go as a whole. 
\item[*]{\bf Does the wave have a preferred direction? (Clockwise or counter-clockwise?)}
\\
clockwise 
\item[*]{\bf Have you ever seen two waves initiated by a single source? (i.e., the wave leaving in both directions) If yes, please give details, e.g., when and where.}
\\
never 
\item[*]{\bf Have you ever taken part in initiating a wave? What did you do? How many of you were needed? Did you try to influence the direction of the wave from the very beginning?}
\\
I didnt. 
\item[*]{\bf Please, select your geographical location.}
\\
Europe 
\end{itemize}
\smallskip
\item[] {\bf \uline{(18) Oct 11, 2004}}\begin{itemize}
\item[*]{\bf Have you ever taken part in a Mexican wave?}
\\
 YES 
\item[*]{\bf If yes, did you follow your neighbours or you considered rather the motion of the wave as a whole? Please, explain below.}
\\
It is hard to know for sure, but I think I was paying attention to the wave as a whole until it started approaching at which point I started paying more attention to my neighbors 
\item[*]{\bf Does the wave have a preferred direction? (Clockwise or counter-clockwise?)}
\\
clockwise 
\item[*]{\bf Have you ever seen two waves initiated by a single source? (i.e., the wave leaving in both directions) If yes, please give details, e.g., when and where.}
\\
yes, but both waves moved in the same direction. I have seen multiple waves started by the same source, again moving in one direction. University of Florida, late 80s early 90s. 
\item[*]{\bf Have you ever taken part in initiating a wave? What did you do? How many of you were needed? Did you try to influence the direction of the wave from the very beginning?}
\\
Yes. We had a handful of friends scream and stand up with arms in air. It always went clockwise. I am curious, does the model consider nonreactors, people who will not participate despite the fact everyone else is. Or does everyone have the same threshold of excitation? 
\item[*]{\bf Please, select your geographical location.}
\\
North America 
\end{itemize}
\smallskip
\item[] {\bf \uline{(19) Oct 19, 2004}}\begin{itemize}
\item[*]{\bf Have you ever taken part in a Mexican wave?}
\\
 YES 
\item[*]{\bf If yes, did you follow your neighbours or you considered rather the motion of the wave as a whole? Please, explain below.}
\\
both. i stand up with my arms raised when my neighbor stands up, but i also consider the waves motion as a whole. 
\item[*]{\bf Does the wave have a preferred direction? (Clockwise or counter-clockwise?)}
\\
clockwise 
\item[*]{\bf Have you ever seen two waves initiated by a single source? (i.e., the wave leaving in both directions) If yes, please give details, e.g., when and where.}
\\
no 
\item[*]{\bf Have you ever taken part in initiating a wave? What did you do? How many of you were needed? Did you try to influence the direction of the wave from the very beginning?}
\\
no 
\item[*]{\bf Please, select your geographical location.}
\\
North America 
\end{itemize}
\smallskip
\item[] {\bf \uline{(20) Nov 19, 2004}}\begin{itemize}
\item[*]{\bf Have you ever taken part in a Mexican wave?}
\\
 YES 
\item[*]{\bf If yes, did you follow your neighbours or you considered rather the motion of the wave as a whole? Please, explain below.}
\\
Both, when paying attention to the wave I followed its motion and anticapted its arrival. When watching the game I joined as my neighbors did. 
\item[*]{\bf Does the wave have a preferred direction? (Clockwise or counter-clockwise?)}
\\
clockwise 
\item[*]{\bf Have you ever seen two waves initiated by a single source? (i.e., the wave leaving in both directions) If yes, please give details, e.g., when and where.}
\\
Yes. At a Univeristy of Michigan footbal game, originating in the student section. 
\item[*]{\bf Have you ever taken part in initiating a wave? What did you do? How many of you were needed? Did you try to influence the direction of the wave from the very beginning?}
\\
No. 
\item[*]{\bf Please, select your geographical location.}
\\
North America 
\end{itemize}

\subsubsection*{Answers 21 to 30 in the online survey}
\medskip
\smallskip
\item[] {\bf \uline{(21) Nov 20, 2004}}\begin{itemize}
\item[*]{\bf Have you ever taken part in a Mexican wave?}
\\
 YES 
\item[*]{\bf If yes, did you follow your neighbours or you considered rather the motion of the wave as a whole? Please, explain below.}
\\
I tend to be behind the first time the wave comes, activating as my neighbors go from active to refracter, then time with the wave as a whole after that(as its the wave as a whole thats entertaining). 
\item[*]{\bf Does the wave have a preferred direction? (Clockwise or counter-clockwise?)}
\\
Perhaps... I find it strangely hard to imagine a right-to-left wave, but that means I imagine it clockwise when opposite me, and counter-clockwise when I participate. 
\item[*]{\bf Have you ever seen two waves initiated by a single source? (i.e., the wave leaving in both directions) If yes, please give details, e.g., when and where.}
\\
At the San Diego Jack Murphy Stadium when I was a little kid(~15yr ago?), I saw one with a counter-wave that moved very slowly for a while before being swept up by the faster wave. 
\item[*]{\bf Have you ever taken part in initiating a wave? What did you do? How many of you were needed? Did you try to influence the direction of the wave from the very beginning?}
\\
No. 
\item[*]{\bf Please, select your geographical location.}
\\
North America 
\end{itemize}
\smallskip
\item[] {\bf \uline{(22) Jan 3, 2005}}\begin{itemize}
\item[*]{\bf Have you ever taken part in a Mexican wave?}
\\
 YES 
\item[*]{\bf If yes, did you follow your neighbours or you considered rather the motion of the wave as a whole? Please, explain below.}
\\
The whole wave (I think). We have a problem in that the MCG is being renovated, so you have to imagine the wave travelling through the part of the stand that is still under construction. 
\item[*]{\bf Does the wave have a preferred direction? (Clockwise or counter-clockwise?)}
\\
In Melbourne (cricket games at the MCG in particular) anticlockwise 
\item[*]{\bf Have you ever seen two waves initiated by a single source? (i.e., the wave leaving in both directions) If yes, please give details, e.g., when and where.}
\\
No, one section of crowd will count down the start of the wave distinctly pointing in the direction intended, emphasising a point with each number, ten (point), nine (point), eight, (point), etc. 
\item[*]{\bf Have you ever taken part in initiating a wave? What did you do? How many of you were needed? Did you try to influence the direction of the wave from the very beginning?}
\\
--
\item[*]{\bf Please, select your geographical location.}
\\
Australia 
\end{itemize}
\smallskip
\item[] {\bf \uline{(23) Jan 13, 2005}}\begin{itemize}
\item[*]{\bf Have you ever taken part in a Mexican wave?}
\\
 YES 
\item[*]{\bf If yes, did you follow your neighbours or you considered rather the motion of the wave as a whole? Please, explain below.}
\\
i think i would consider the wave as a whole. 
\item[*]{\bf Does the wave have a preferred direction? (Clockwise or counter-clockwise?)}
\\
the wave knows no direction 
\item[*]{\bf Have you ever seen two waves initiated by a single source? (i.e., the wave leaving in both directions) If yes, please give details, e.g., when and where.}
\\
no cant say that i have sorry 
\item[*]{\bf Have you ever taken part in initiating a wave? What did you do? How many of you were needed? Did you try to influence the direction of the wave from the very beginning?}
\\
cant say that i have started a wave sorry 
\item[*]{\bf Please, select your geographical location.}
\\
North America 
\end{itemize}
\smallskip
\item[] {\bf \uline{(24) Jan 16, 2005}}\begin{itemize}
\item[*]{\bf Have you ever taken part in a Mexican wave?}
\\
 YES 
\item[*]{\bf If yes, did you follow your neighbours or you considered rather the motion of the wave as a whole? Please, explain below.}
\\
I watched the wave moving around the stadium until it reaches my section and take cues from my neighbours as to when to jump up. 
\item[*]{\bf Does the wave have a preferred direction? (Clockwise or counter-clockwise?)}
\\
As far as I know it is always counter-clockwise looking from above 
\item[*]{\bf Have you ever seen two waves initiated by a single source? (i.e., the wave leaving in both directions) If yes, please give details, e.g., when and where.}
\\
No 
\item[*]{\bf Have you ever taken part in initiating a wave? What did you do? How many of you were needed? Did you try to influence the direction of the wave from the very beginning?}
\\
I have, we needed probably 50 otherwise it would fail, I never tried to influence the direction however others may have. I am very interested in your opinion as to how direction is decided. 
\item[*]{\bf Please, select your geographical location.}
\\
Australia 
\end{itemize}
\smallskip
\item[] {\bf \uline{(25) Jan 16, 2005}}\begin{itemize}
\item[*]{\bf Have you ever taken part in a Mexican wave?}
\\
 YES 
\item[*]{\bf If yes, did you follow your neighbours or you considered rather the motion of the wave as a whole? Please, explain below.}
\\
I watch the wave move towards me, attempting to time my top of my rise with what i determined was the middle of the wave 
\item[*]{\bf Does the wave have a preferred direction? (Clockwise or counter-clockwise?)}
\\
Counter-clockwise 
\item[*]{\bf Have you ever seen two waves initiated by a single source? (i.e., the wave leaving in both directions) If yes, please give details, e.g., when and where.}
\\
No, all that i have seen go in the counterclockwise direction 
\item[*]{\bf Have you ever taken part in initiating a wave? What did you do? How many of you were needed? Did you try to influence the direction of the wave from the very beginning?}
\\
I think i have but cant remember any details. 
\item[*]{\bf Please, select your geographical location.}
\\
Australia 
\end{itemize}
\smallskip
\item[] {\bf \uline{(26) Feb 2, 2005}}\begin{itemize}
\item[*]{\bf Have you ever taken part in a Mexican wave?}
\\
 YES 
\item[*]{\bf If yes, did you follow your neighbours or you considered rather the motion of the wave as a whole? Please, explain below.}
\\
Considered the motion of the wave as a whole - I always see them coming 
\item[*]{\bf Does the wave have a preferred direction? (Clockwise or counter-clockwise?)}
\\
counter-clockwise 
\item[*]{\bf Have you ever seen two waves initiated by a single source? (i.e., the wave leaving in both directions) If yes, please give details, e.g., when and where.}
\\
No 
\item[*]{\bf Have you ever taken part in initiating a wave? What did you do? How many of you were needed? Did you try to influence the direction of the wave from the very beginning?}
\\
No 
\item[*]{\bf Please, select your geographical location.}
\\
Australia 
\end{itemize}
\smallskip
\item[] {\bf \uline{(27) Feb 14, 2005}}\begin{itemize}
\item[*]{\bf Have you ever taken part in a Mexican wave?}
\\
 YES 
\item[*]{\bf If yes, did you follow your neighbours or you considered rather the motion of the wave as a whole? Please, explain below.}
\\
1. I watched the wave, 2. when the wave was close I followed the neighbours 
\item[*]{\bf Does the wave have a preferred direction? (Clockwise or counter-clockwise?)}
\\
counter 
\item[*]{\bf Have you ever seen two waves initiated by a single source? (i.e., the wave leaving in both directions) If yes, please give details, e.g., when and where.}
\\
if two waves start from the same source, people decide to follow the stronger wave 
\item[*]{\bf Have you ever taken part in initiating a wave? What did you do? How many of you were needed? Did you try to influence the direction of the wave from the very beginning?}
\\
I like to go to Mexican soccer, and my believe is that the waves are not causal. Probably, if you talk in front of 100 or 50 and you convince them to make a wave you coul do 
\item[*]{\bf Please, select your geographical location.}
\\
North America 
\end{itemize}
\smallskip
\item[] {\bf \uline{(28) Mar 4, 2005}}\begin{itemize}
\item[*]{\bf Have you ever taken part in a Mexican wave?}
\\
 YES 
\item[*]{\bf If yes, did you follow your neighbours or you considered rather the motion of the wave as a whole? Please, explain below.}
\\
I followed the wave as a whole rather than my neighbors. I paced the wave and jumped up at the time the wave hit. 
\item[*]{\bf Does the wave have a preferred direction? (Clockwise or counter-clockwise?)}
\\
All waves I have been in have been clockwise, right to left. 
\item[*]{\bf Have you ever seen two waves initiated by a single source? (i.e., the wave leaving in both directions) If yes, please give details, e.g., when and where.}
\\
I have never seen a wave split into two directions. 
\item[*]{\bf Have you ever taken part in initiating a wave? What did you do? How many of you were needed? Did you try to influence the direction of the wave from the very beginning?}
\\
I initiated a wave with a group of about ten friends at an Oakland Athletics baseball game in Oakland, California, USA in 2004. The ten of us attempted to start a wave twice before strangers in our local area noticed and joined in the initiation. By the fourth time the Wave was successfully launched to our joy. 
\item[*]{\bf Please, select your geographical location.}
\\
North America 
\end{itemize}
\smallskip
\item[] {\bf \uline{(29) Mar 14, 2005}}\begin{itemize}
\item[*]{\bf Have you ever taken part in a Mexican wave?}
\\
 YES 
\item[*]{\bf If yes, did you follow your neighbours or you considered rather the motion of the wave as a whole? Please, explain below.}
\\
the wave as a whole 
\item[*]{\bf Does the wave have a preferred direction? (Clockwise or counter-clockwise?)}
\\
No its dependent on where the initiator, (i.e. the person lively enough to want to start the wave), sits in relation to his friends. Since it cant be started by one person, not notifying a few people in row beforehand, the intiator will more than likely get the people he knows at the furthest end to where hes sitting start it. Therefore it is dependent on the intiators position among his friends. 
\item[*]{\bf Have you ever seen two waves initiated by a single source? (i.e., the wave leaving in both directions) If yes, please give details, e.g., when and where.}
\\
No but thats a great idea 
\item[*]{\bf Have you ever taken part in initiating a wave? What did you do? How many of you were needed? Did you try to influence the direction of the wave from the very beginning?}
\\
Yes on a trip to Yankee Stadium New York(although Im Irish).I was there in a big group of international students I got on well with. Proudly it was my idea to start it and yes the direction was pretty much set out since we got a few of us in a row to do it. I remember we did it a few times since the first didnt get a reaction. The second time I think there was one random guy who did it and we cheered him. Then since I presume people around us were catching on to what we were doing the third time it went to the end of the stadium. Now I say end of the stadium because Yankee Stadium(a base ball stadium) is a crest shape i.e. its not a complete circle. Regardless I think a lot of people were starting to see the little waves and a lasting wave was started (sadly not by myself but somewhere else). This wave when it reached the side of the stadium the other side kept it going and around and around it went. I still feel I laid the ground work for this lasting wave though. 
\item[*]{\bf Please, select your geographical location.}
\\
Europe 
\end{itemize}
\smallskip
\item[] {\bf \uline{(30) Mar 22, 2005}}\begin{itemize}
\item[*]{\bf Have you ever taken part in a Mexican wave?}
\\
 YES 
\item[*]{\bf If yes, did you follow your neighbours or you considered rather the motion of the wave as a whole? Please, explain below.}
\\
we follow the neighbours, when he goes up ill start to raise 
\item[*]{\bf Does the wave have a preferred direction? (Clockwise or counter-clockwise?)}
\\
I never paid attention, but i think it comes from left 
\item[*]{\bf Have you ever seen two waves initiated by a single source? (i.e., the wave leaving in both directions) If yes, please give details, e.g., when and where.}
\\
never 
\item[*]{\bf Have you ever taken part in initiating a wave? What did you do? How many of you were needed? Did you try to influence the direction of the wave from the very beginning?}
\\
never 
\item[*]{\bf Please, select your geographical location.}
\\
North America 
\end{itemize}

\subsubsection*{Answers 31 to 40 in the online survey}
\medskip
\smallskip
\item[] {\bf \uline{(31) Apr 1, 2005}}\begin{itemize}
\item[*]{\bf Have you ever taken part in a Mexican wave?}
\\
 YES 
\item[*]{\bf If yes, did you follow your neighbours or you considered rather the motion of the wave as a whole? Please, explain below.}
\\
You see when it is in front, and have an idea of when its coming, but when its close you cant see it, because the neighbours are in the way, so you stand up after they do. 
\item[*]{\bf Does the wave have a preferred direction? (Clockwise or counter-clockwise?)}
\\
Comes from your right and leaves to your left. So from the top of the stadium, clockwise 
\item[*]{\bf Have you ever seen two waves initiated by a single source? (i.e., the wave leaving in both directions) If yes, please give details, e.g., when and where.}
\\
no. Everybody knows the direction. Maybe in begginning one the people near where it starts also rise, even if they are on the other side, so it starts big, but then its clockwise. 
\item[*]{\bf Have you ever taken part in initiating a wave? What did you do? How many of you were needed? Did you try to influence the direction of the wave from the very beginning?}
\\
No, but Ive seen people try in places with not so much public, and separated, and it doesnt work 
\item[*]{\bf Please, select your geographical location.}
\\
North America 
\end{itemize}
\smallskip
\item[] {\bf \uline{(32) Apr 1, 2005}}\begin{itemize}
\item[*]{\bf Have you ever taken part in a Mexican wave?}
\\
 YES 
\item[*]{\bf If yes, did you follow your neighbours or you considered rather the motion of the wave as a whole? Please, explain below.}
\\
one more thing, sorry. It should be written separately: la ola (not laola, as it is in the page). ola is wave (of the sea, not any wave), so la ola is the wave 
\item[*]{\bf Does the wave have a preferred direction? (Clockwise or counter-clockwise?)}
\\
--
\item[*]{\bf Have you ever seen two waves initiated by a single source? (i.e., the wave leaving in both directions) If yes, please give details, e.g., when and where.}
\\
--
\item[*]{\bf Have you ever taken part in initiating a wave? What did you do? How many of you were needed? Did you try to influence the direction of the wave from the very beginning?}
\\
--
\item[*]{\bf Please, select your geographical location.}
\\
--
\end{itemize}
\smallskip
\item[] {\bf \uline{(33) Apr 14, 2005}}\begin{itemize}
\item[*]{\bf Have you ever taken part in a Mexican wave?}
\\
 YES 
\item[*]{\bf If yes, did you follow your neighbours or you considered rather the motion of the wave as a whole? Please, explain below.}
\\
It was done because everyone was doing it. It wasnt rehearsed. Therefore, it didnt look very pretty from our view. Yet, it was completed with dignity at a High School Basketball game. Deb 
\item[*]{\bf Does the wave have a preferred direction? (Clockwise or counter-clockwise?)}
\\
to the left 
\item[*]{\bf Have you ever seen two waves initiated by a single source? (i.e., the wave leaving in both directions) If yes, please give details, e.g., when and where.}
\\
No 
\item[*]{\bf Have you ever taken part in initiating a wave? What did you do? How many of you were needed? Did you try to influence the direction of the wave from the very beginning?}
\\
No 
\item[*]{\bf Please, select your geographical location.}
\\
North America 
\end{itemize}
\smallskip
\item[] {\bf \uline{(34) Apr 14, 2005}}\begin{itemize}
\item[*]{\bf Have you ever taken part in a Mexican wave?}
\\
 YES 
\item[*]{\bf If yes, did you follow your neighbours or you considered rather the motion of the wave as a whole? Please, explain below.}
\\
Neighbors--You cant see the whole when youre in the active part of the wave. 
\item[*]{\bf Does the wave have a preferred direction? (Clockwise or counter-clockwise?)}
\\
Ive never seen a counter-clockwise wave! 
\item[*]{\bf Have you ever seen two waves initiated by a single source? (i.e., the wave leaving in both directions) If yes, please give details, e.g., when and where.}
\\
nope 
\item[*]{\bf Have you ever taken part in initiating a wave? What did you do? How many of you were needed? Did you try to influence the direction of the wave from the very beginning?}
\\
nope 
\item[*]{\bf Please, select your geographical location.}
\\
North America 
\end{itemize}
\smallskip
\item[] {\bf \uline{(35) Apr 14, 2005}}\begin{itemize}
\item[*]{\bf Have you ever taken part in a Mexican wave?}
\\
 YES 
\item[*]{\bf If yes, did you follow your neighbours or you considered rather the motion of the wave as a whole? Please, explain below.}
\\
Both at different times, but I find it easier to consider it as a wave rather than a collection of moving particles (people). 
\item[*]{\bf Does the wave have a preferred direction? (Clockwise or counter-clockwise?)}
\\
Clockwise. 
\item[*]{\bf Have you ever seen two waves initiated by a single source? (i.e., the wave leaving in both directions) If yes, please give details, e.g., when and where.}
\\
No. 
\item[*]{\bf Have you ever taken part in initiating a wave? What did you do? How many of you were needed? Did you try to influence the direction of the wave from the very beginning?}
\\
In a medium-sized stadium, I have been able to start a wave with as few as one other person. We werent trying to go a certain direction, but we must have influenced it because we were next to each other and we stood up at separated times. It only seems to work if something has happened in the game that causes happiness in most of the spectators. 
\item[*]{\bf Please, select your geographical location.}
\\
North America 
\end{itemize}
\smallskip
\item[] {\bf \uline{(36) Apr 14, 2005}}\begin{itemize}
\item[*]{\bf Have you ever taken part in a Mexican wave?}
\\
 YES 
\item[*]{\bf If yes, did you follow your neighbours or you considered rather the motion of the wave as a whole? Please, explain below.}
\\
Followed my neighbors. 
\item[*]{\bf Does the wave have a preferred direction? (Clockwise or counter-clockwise?)}
\\
Clockwise 
\item[*]{\bf Have you ever seen two waves initiated by a single source? (i.e., the wave leaving in both directions) If yes, please give details, e.g., when and where.}
\\
No 
\item[*]{\bf Have you ever taken part in initiating a wave? What did you do? How many of you were needed? Did you try to influence the direction of the wave from the very beginning?}
\\
No 
\item[*]{\bf Please, select your geographical location.}
\\
North America 
\end{itemize}
\smallskip
\item[] {\bf \uline{(37) Apr 14, 2005}}\begin{itemize}
\item[*]{\bf Have you ever taken part in a Mexican wave?}
\\
 YES 
\item[*]{\bf If yes, did you follow your neighbours or you considered rather the motion of the wave as a whole? Please, explain below.}
\\
I followed the motion of the wave as a whole. 
\item[*]{\bf Does the wave have a preferred direction? (Clockwise or counter-clockwise?)}
\\
Clockwise 
\item[*]{\bf Have you ever seen two waves initiated by a single source? (i.e., the wave leaving in both directions) If yes, please give details, e.g., when and where.}
\\
Yes. At the University of Michigan in Ann Arbor I experienced waves like none other I have experienced. It always starts in a clockwise direction and at one point, through coordinating hand signals, the wave will change. There are four potential changes. One is to speed up the wave such that it is nothing more than a quick flash of the hands in the air. The second is a slow down of the wave such that it appears it is moving in slow motion. The third is to reverse the direction of the wave, and the fourth is to split the wave into two opposite-moving directions. This type of wave can be witnessed at almost any University of Michigan football game. The coordinators of the changes are an informal group of students located at the front of the student section of the 110,000-person football stadium. 
\item[*]{\bf Have you ever taken part in initiating a wave? What did you do? How many of you were needed? Did you try to influence the direction of the wave from the very beginning?}
\\
No, I have not taken part in initiating a wave. 
\item[*]{\bf Please, select your geographical location.}
\\
North America 
\end{itemize}
\smallskip
\item[] {\bf \uline{(38) Apr 14, 2005}}\begin{itemize}
\item[*]{\bf Have you ever taken part in a Mexican wave?}
\\
 YES 
\item[*]{\bf If yes, did you follow your neighbours or you considered rather the motion of the wave as a whole? Please, explain below.}
\\
Crowd phenomenon - all postitive. I LOVE the wave! Wave as a whole. 
\item[*]{\bf Does the wave have a preferred direction? (Clockwise or counter-clockwise?)}
\\
clockwise 
\item[*]{\bf Have you ever seen two waves initiated by a single source? (i.e., the wave leaving in both directions) If yes, please give details, e.g., when and where.}
\\
nope 
\item[*]{\bf Have you ever taken part in initiating a wave? What did you do? How many of you were needed? Did you try to influence the direction of the wave from the very beginning?}
\\
Yes, with teen groups. No directional influence 
\item[*]{\bf Please, select your geographical location.}
\\
North America 
\end{itemize}
\smallskip
\item[] {\bf \uline{(39) Apr 14, 2005}}\begin{itemize}
\item[*]{\bf Have you ever taken part in a Mexican wave?}
\\
 YES 
\item[*]{\bf If yes, did you follow your neighbours or you considered rather the motion of the wave as a whole? Please, explain below.}
\\
I watch the wave as a whole. It seems to take on a life of its own. 
\item[*]{\bf Does the wave have a preferred direction? (Clockwise or counter-clockwise?)}
\\
I seem to remember them going clockwise, most of the time. 
\item[*]{\bf Have you ever seen two waves initiated by a single source? (i.e., the wave leaving in both directions) If yes, please give details, e.g., when and where.}
\\
No. 
\item[*]{\bf Have you ever taken part in initiating a wave? What did you do? How many of you were needed? Did you try to influence the direction of the wave from the very beginning?}
\\
Yes, my friends and I tried to initiate a wave once. Six of us were sitting together, but we couldnt have done it without some nearby wave enthusiasts who noticed right away what we were doing. They helped us out, and timed their cheers with ours.It took several attempts before we got the attention of enough people! I guess there were about a dozen of us at the very beginning. We finally did end up trying to influence the direction of the wave - one of my friends ended up getting out of his seat, and ran down the aisle yelling and waving his arms. Then he quietly went back to where he started, and ran back the other way again, yelling and waving. The wave started to build up a few times, but always died quickly. Suddenly, it seemed to hit critical mass - enough people had noticed what we were trying to do, and our wave rounded the stadium three or four times. 
\item[*]{\bf Please, select your geographical location.}
\\
North America 
\end{itemize}
\smallskip
\item[] {\bf \uline{(40) Apr 14, 2005}}\begin{itemize}
\item[*]{\bf Have you ever taken part in a Mexican wave?}
\\
 YES 
\item[*]{\bf If yes, did you follow your neighbours or you considered rather the motion of the wave as a whole? Please, explain below.}
\\
followed neighbours 
\item[*]{\bf Does the wave have a preferred direction? (Clockwise or counter-clockwise?)}
\\
clockwise 
\item[*]{\bf Have you ever seen two waves initiated by a single source? (i.e., the wave leaving in both directions) If yes, please give details, e.g., when and where.}
\\
No 
\item[*]{\bf Have you ever taken part in initiating a wave? What did you do? How many of you were needed? Did you try to influence the direction of the wave from the very beginning?}
\\
No 
\item[*]{\bf Please, select your geographical location.}
\\
North America 
\end{itemize}

\subsubsection*{Answers 41 to 50 in the online survey}
\medskip
\smallskip
\item[] {\bf \uline{(41) Apr 14, 2005}}\begin{itemize}
\item[*]{\bf Have you ever taken part in a Mexican wave?}
\\
 YES 
\item[*]{\bf If yes, did you follow your neighbours or you considered rather the motion of the wave as a whole? Please, explain below.}
\\
Followed neighbors. But I have watched the wave propogate around and try to anticipate the arrival so I could try to speed up the wave a little bit (no success). 
\item[*]{\bf Does the wave have a preferred direction? (Clockwise or counter-clockwise?)}
\\
Counterclockwise (wave arrives from my left) 
\item[*]{\bf Have you ever seen two waves initiated by a single source? (i.e., the wave leaving in both directions) If yes, please give details, e.g., when and where.}
\\
No 
\item[*]{\bf Have you ever taken part in initiating a wave? What did you do? How many of you were needed? Did you try to influence the direction of the wave from the very beginning?}
\\
No 
\item[*]{\bf Please, select your geographical location.}
\\
North America 
\end{itemize}
\smallskip
\item[] {\bf \uline{(42) Apr 14, 2005}}\begin{itemize}
\item[*]{\bf Have you ever taken part in a Mexican wave?}
\\
 YES 
\item[*]{\bf If yes, did you follow your neighbours or you considered rather the motion of the wave as a whole? Please, explain below.}
\\
I considered the motion fo the wave as a whole. If you follow your neighbors your reaction is too slow thereby putting you off the proper wave course. It is hence necessary to judge for yourself the precise moment to engage in the wave. 
\item[*]{\bf Does the wave have a preferred direction? (Clockwise or counter-clockwise?)}
\\
clockwise 
\item[*]{\bf Have you ever seen two waves initiated by a single source? (i.e., the wave leaving in both directions) If yes, please give details, e.g., when and where.}
\\
No 
\item[*]{\bf Have you ever taken part in initiating a wave? What did you do? How many of you were needed? Did you try to influence the direction of the wave from the very beginning?}
\\
Yes. There were approximately ten of us. Who were yelling do the wave. W then began a countdown from three and indicated to those in teh seats above us to wave. 
\item[*]{\bf Please, select your geographical location.}
\\
North America 
\end{itemize}
\smallskip
\item[] {\bf \uline{(43) Apr 14, 2005}}\begin{itemize}
\item[*]{\bf Have you ever taken part in a Mexican wave?}
\\
 YES 
\item[*]{\bf If yes, did you follow your neighbours or you considered rather the motion of the wave as a whole? Please, explain below.}
\\
Followed neighbors. 
\item[*]{\bf Does the wave have a preferred direction? (Clockwise or counter-clockwise?)}
\\
Clockwaves 
\item[*]{\bf Have you ever seen two waves initiated by a single source? (i.e., the wave leaving in both directions) If yes, please give details, e.g., when and where.}
\\
No 
\item[*]{\bf Have you ever taken part in initiating a wave? What did you do? How many of you were needed? Did you try to influence the direction of the wave from the very beginning?}
\\
No 
\item[*]{\bf Please, select your geographical location.}
\\
North America 
\end{itemize}
\smallskip
\item[] {\bf \uline{(44) Apr 14, 2005}}\begin{itemize}
\item[*]{\bf Have you ever taken part in a Mexican wave?}
\\
 YES 
\item[*]{\bf If yes, did you follow your neighbours or you considered rather the motion of the wave as a whole? Please, explain below.}
\\
followed my neighbors 
\item[*]{\bf Does the wave have a preferred direction? (Clockwise or counter-clockwise?)}
\\
clockwise 
\item[*]{\bf Have you ever seen two waves initiated by a single source? (i.e., the wave leaving in both directions) If yes, please give details, e.g., when and where.}
\\
No 
\item[*]{\bf Have you ever taken part in initiating a wave? What did you do? How many of you were needed? Did you try to influence the direction of the wave from the very beginning?}
\\
Yes, about three of my friends and I started standing up and raising our arms above our head until several people around us started and it took off - no we did not try to influence the direction however, the waves we intiated always went clockwise. 
\item[*]{\bf Please, select your geographical location.}
\\
North America 
\end{itemize}
\smallskip
\item[] {\bf \uline{(45) Apr 14, 2005}}\begin{itemize}
\item[*]{\bf Have you ever taken part in a Mexican wave?}
\\
 YES 
\item[*]{\bf If yes, did you follow your neighbours or you considered rather the motion of the wave as a whole? Please, explain below.}
\\
GENERAL WAVE AS A WHOLE 
\item[*]{\bf Does the wave have a preferred direction? (Clockwise or counter-clockwise?)}
\\
 CLOCKWISE 
\item[*]{\bf Have you ever seen two waves initiated by a single source? (i.e., the wave leaving in both directions) If yes, please give details, e.g., when and where.}
\\
NO 
\item[*]{\bf Have you ever taken part in initiating a wave? What did you do? How many of you were needed? Did you try to influence the direction of the wave from the very beginning?}
\\
YES. ME AND 5 FRIENDS GOT A SECTION AND SAID WERRE STARTING THE WAVE. ON THRE 1.. 2... 3! YES 
\item[*]{\bf Please, select your geographical location.}
\\
North America 
\end{itemize}
\smallskip
\item[] {\bf \uline{(46) Apr 14, 2005}}\begin{itemize}
\item[*]{\bf Have you ever taken part in a Mexican wave?}
\\
 YES 
\item[*]{\bf If yes, did you follow your neighbours or you considered rather the motion of the wave as a whole? Please, explain below.}
\\
wave as a whole 
\item[*]{\bf Does the wave have a preferred direction? (Clockwise or counter-clockwise?)}
\\
counter-clockwise 
\item[*]{\bf Have you ever seen two waves initiated by a single source? (i.e., the wave leaving in both directions) If yes, please give details, e.g., when and where.}
\\
No 
\item[*]{\bf Have you ever taken part in initiating a wave? What did you do? How many of you were needed? Did you try to influence the direction of the wave from the very beginning?}
\\
No 
\item[*]{\bf Please, select your geographical location.}
\\
North America 
\end{itemize}
\smallskip
\item[] {\bf \uline{(47) Apr 14, 2005}}\begin{itemize}
\item[*]{\bf Have you ever taken part in a Mexican wave?}
\\
 YES 
\item[*]{\bf If yes, did you follow your neighbours or you considered rather the motion of the wave as a whole? Please, explain below.}
\\
Followed my neighbors. 
\item[*]{\bf Does the wave have a preferred direction? (Clockwise or counter-clockwise?)}
\\
Ive been in many that have traveled in both directions, usually counter-clockwise to clockwise. 
\item[*]{\bf Have you ever seen two waves initiated by a single source? (i.e., the wave leaving in both directions) If yes, please give details, e.g., when and where.}
\\
No. 
\item[*]{\bf Have you ever taken part in initiating a wave? What did you do? How many of you were needed? Did you try to influence the direction of the wave from the very beginning?}
\\
No I havent. 
\item[*]{\bf Please, select your geographical location.}
\\
North America 
\end{itemize}
\smallskip
\item[] {\bf \uline{(48) Apr 14, 2005}}\begin{itemize}
\item[*]{\bf Have you ever taken part in a Mexican wave?}
\\
 YES 
\item[*]{\bf If yes, did you follow your neighbours or you considered rather the motion of the wave as a whole? Please, explain below.}
\\
I think I followed my neighbours. 
\item[*]{\bf Does the wave have a preferred direction? (Clockwise or counter-clockwise?)}
\\
clockwise 
\item[*]{\bf Have you ever seen two waves initiated by a single source? (i.e., the wave leaving in both directions) If yes, please give details, e.g., when and where.}
\\
no 
\item[*]{\bf Have you ever taken part in initiating a wave? What did you do? How many of you were needed? Did you try to influence the direction of the wave from the very beginning?}
\\
no. 
\item[*]{\bf Please, select your geographical location.}
\\
North America 
\end{itemize}
\smallskip
\item[] {\bf \uline{(49) Apr 14, 2005}}\begin{itemize}
\item[*]{\bf Have you ever taken part in a Mexican wave?}
\\
 YES 
\item[*]{\bf If yes, did you follow your neighbours or you considered rather the motion of the wave as a whole? Please, explain below.}
\\
If paying attention, I considered the waves motion. If not paying attention, I was alerted to the wave by the action of neighbours 
\item[*]{\bf Does the wave have a preferred direction? (Clockwise or counter-clockwise?)}
\\
All the ones Ive seen have have been clockwise 
\item[*]{\bf Have you ever seen two waves initiated by a single source? (i.e., the wave leaving in both directions) If yes, please give details, e.g., when and where.}
\\
no 
\item[*]{\bf Have you ever taken part in initiating a wave? What did you do? How many of you were needed? Did you try to influence the direction of the wave from the very beginning?}
\\
Always started by counting down slowly - and as loudly as possible - from 5. First two or three people (usually schoolboys or fairly drunken fans!), or a larger group of friends, would start counting, then others would join in as the count got down to 1. Successful waves started with a roar. 
\item[*]{\bf Please, select your geographical location.}
\\
Europe 
\end{itemize}
\smallskip
\item[] {\bf \uline{(50) Apr 15, 2005}}\begin{itemize}
\item[*]{\bf Have you ever taken part in a Mexican wave?}
\\
 YES 
\item[*]{\bf If yes, did you follow your neighbours or you considered rather the motion of the wave as a whole? Please, explain below.}
\\
I followed my neighbors. 
\item[*]{\bf Does the wave have a preferred direction? (Clockwise or counter-clockwise?)}
\\
Definitely from left to right (counterclockwise) in smaller stadiums, however, in larger stadiums, it tends to be started on both ends and go in to the center, or vice versa. 
\item[*]{\bf Have you ever seen two waves initiated by a single source? (i.e., the wave leaving in both directions) If yes, please give details, e.g., when and where.}
\\
Yes, at a Colorado Rockies game, by a large group of people all together starting it up right behind home plate. 
\item[*]{\bf Have you ever taken part in initiating a wave? What did you do? How many of you were needed? Did you try to influence the direction of the wave from the very beginning?}
\\
I have not taken part in initiating a wave. One thing you might want to know, however, is that laola doesnt mean anything in spanish, its actually la ola, which refers to a singular wave. 
\item[*]{\bf Please, select your geographical location.}
\\
North America 
\end{itemize}

\subsubsection*{Answers 51 to 60 in the online survey}
\medskip
\smallskip
\item[] {\bf \uline{(51) Apr 15, 2005}}\begin{itemize}
\item[*]{\bf Have you ever taken part in a Mexican wave?}
\\
 YES 
\item[*]{\bf If yes, did you follow your neighbours or you considered rather the motion of the wave as a whole? Please, explain below.}
\\
I tend to follow the motion of the wave - to keep the entire big picture. 
\item[*]{\bf Does the wave have a preferred direction? (Clockwise or counter-clockwise?)}
\\
Either - just keep it going! 
\item[*]{\bf Have you ever seen two waves initiated by a single source? (i.e., the wave leaving in both directions) If yes, please give details, e.g., when and where.}
\\
I have not experienced it. 
\item[*]{\bf Have you ever taken part in initiating a wave? What did you do? How many of you were needed? Did you try to influence the direction of the wave from the very beginning?}
\\
I have - usually jumping up saying Lets do the wave. 
\item[*]{\bf Please, select your geographical location.}
\\
North America 
\end{itemize}
\smallskip
\item[] {\bf \uline{(52) Apr 15, 2005}}\begin{itemize}
\item[*]{\bf Have you ever taken part in a Mexican wave?}
\\
 YES 
\item[*]{\bf If yes, did you follow your neighbours or you considered rather the motion of the wave as a whole? Please, explain below.}
\\
I considered the motion of the wave as a whole, i saw it roll along the stadium like a real wave. 
\item[*]{\bf Does the wave have a preferred direction? (Clockwise or counter-clockwise?)}
\\
clockwise 
\item[*]{\bf Have you ever seen two waves initiated by a single source? (i.e., the wave leaving in both directions) If yes, please give details, e.g., when and where.}
\\
no 
\item[*]{\bf Have you ever taken part in initiating a wave? What did you do? How many of you were needed? Did you try to influence the direction of the wave from the very beginning?}
\\
yes, me and a group of friends stood up and yelled wave!!!! 
\item[*]{\bf Please, select your geographical location.}
\\
North America 
\end{itemize}
\smallskip
\item[] {\bf \uline{(53) Apr 15, 2005}}\begin{itemize}
\item[*]{\bf Have you ever taken part in a Mexican wave?}
\\
 YES 
\item[*]{\bf If yes, did you follow your neighbours or you considered rather the motion of the wave as a whole? Please, explain below.}
\\
the wave as a whole - i like to push it faster 
\item[*]{\bf Does the wave have a preferred direction? (Clockwise or counter-clockwise?)}
\\
clockwise 
\item[*]{\bf Have you ever seen two waves initiated by a single source? (i.e., the wave leaving in both directions) If yes, please give details, e.g., when and where.}
\\
no 
\item[*]{\bf Have you ever taken part in initiating a wave? What did you do? How many of you were needed? Did you try to influence the direction of the wave from the very beginning?}
\\
yes - we had a drum major at a football game run back and forth to let the crowd know what we were doing, and then we (the band) started it, and the drum major ran along with it to keep it going. 
\item[*]{\bf Please, select your geographical location.}
\\
North America 
\end{itemize}
\smallskip
\item[] {\bf \uline{(54) Apr 15, 2005}}\begin{itemize}
\item[*]{\bf Have you ever taken part in a Mexican wave?}
\\
 YES 
\item[*]{\bf If yes, did you follow your neighbours or you considered rather the motion of the wave as a whole? Please, explain below.}
\\
yes - i followed the wave as a whole - and by the way, its La Ola, not laola. 2 separate spanish words meaning the wave 
\item[*]{\bf Does the wave have a preferred direction? (Clockwise or counter-clockwise?)}
\\
clockwise 
\item[*]{\bf Have you ever seen two waves initiated by a single source? (i.e., the wave leaving in both directions) If yes, please give details, e.g., when and where.}
\\
no 
\item[*]{\bf Have you ever taken part in initiating a wave? What did you do? How many of you were needed? Did you try to influence the direction of the wave from the very beginning?}
\\
no 
\item[*]{\bf Please, select your geographical location.}
\\
North America 
\end{itemize}
\smallskip
\item[] {\bf \uline{(55) Apr 15, 2005}}\begin{itemize}
\item[*]{\bf Have you ever taken part in a Mexican wave?}
\\
 YES 
\item[*]{\bf If yes, did you follow your neighbours or you considered rather the motion of the wave as a whole? Please, explain below.}
\\
folowed my neighbors 
\item[*]{\bf Does the wave have a preferred direction? (Clockwise or counter-clockwise?)}
\\
clockwise 
\item[*]{\bf Have you ever seen two waves initiated by a single source? (i.e., the wave leaving in both directions) If yes, please give details, e.g., when and where.}
\\
No 
\item[*]{\bf Have you ever taken part in initiating a wave? What did you do? How many of you were needed? Did you try to influence the direction of the wave from the very beginning?}
\\
Yes, I was part of a group an with a little yelling we got it going 
\item[*]{\bf Please, select your geographical location.}
\\
North America 
\end{itemize}
\smallskip
\item[] {\bf \uline{(56) Apr 16, 2005}}\begin{itemize}
\item[*]{\bf Have you ever taken part in a Mexican wave?}
\\
 YES 
\item[*]{\bf If yes, did you follow your neighbours or you considered rather the motion of the wave as a whole? Please, explain below.}
\\
I am not sure what your question is, but I do not go up exactly with my neighbors but rather purposefully go up just after the person to my right. 
\item[*]{\bf Does the wave have a preferred direction? (Clockwise or counter-clockwise?)}
\\
have only experienced it clock-wise 
\item[*]{\bf Have you ever seen two waves initiated by a single source? (i.e., the wave leaving in both directions) If yes, please give details, e.g., when and where.}
\\
nope 
\item[*]{\bf Have you ever taken part in initiating a wave? What did you do? How many of you were needed? Did you try to influence the direction of the wave from the very beginning?}
\\
yes me and my friend to the left repeatedly jumped up with our arms straight up and made a whoah sound. While I would say there was not a conscious effort to send the wave one way or the other (because I think my friend and I both assumed for whatever reason that it would naturally go clock-wise), I suppose our looking left could have influenced the beginning direction- though alternately, perhaps we only looked left because that is where it was picking up. ??? 
\item[*]{\bf Please, select your geographical location.}
\\
North America 
\end{itemize}
\smallskip
\item[] {\bf \uline{(57) Apr 16, 2005}}\begin{itemize}
\item[*]{\bf Have you ever taken part in a Mexican wave?}
\\
 YES 
\item[*]{\bf If yes, did you follow your neighbours or you considered rather the motion of the wave as a whole? Please, explain below.}
\\
As a whole - the stadium held 80,000+ people and they managed to move together. 
\item[*]{\bf Does the wave have a preferred direction? (Clockwise or counter-clockwise?)}
\\
It seems to go counter-clockwise most often in my experience. But, in the stadium that was just a line it always went clockwise. 
\item[*]{\bf Have you ever seen two waves initiated by a single source? (i.e., the wave leaving in both directions) If yes, please give details, e.g., when and where.}
\\
No 
\item[*]{\bf Have you ever taken part in initiating a wave? What did you do? How many of you were needed? Did you try to influence the direction of the wave from the very beginning?}
\\
Yes, the University of South Carolina Band (300 total members) initated a wave that went around the Williams Brice Stadium at least 8 times. There were about 20-30 that initiated it, but we were in a group. We did start it on one side and it picked up easily because we were in a large group of drunk college students. It was not done by all of the people right when it started though because the people above us could not see that we were initiating it, but they caught it the next time around. The only reason it stopped was because something really important happened in the game and people stopped paying attention. 
\item[*]{\bf Please, select your geographical location.}
\\
North America 
\end{itemize}
\smallskip
\item[] {\bf \uline{(58) Apr 16, 2005}}\begin{itemize}
\item[*]{\bf Have you ever taken part in a Mexican wave?}
\\
 YES 
\item[*]{\bf If yes, did you follow your neighbours or you considered rather the motion of the wave as a whole? Please, explain below.}
\\
I followed my neighbours. 
\item[*]{\bf Does the wave have a preferred direction? (Clockwise or counter-clockwise?)}
\\
Clockwise 
\item[*]{\bf Have you ever seen two waves initiated by a single source? (i.e., the wave leaving in both directions) If yes, please give details, e.g., when and where.}
\\
No 
\item[*]{\bf Have you ever taken part in initiating a wave? What did you do? How many of you were needed? Did you try to influence the direction of the wave from the very beginning?}
\\
Yes, small waves during school events. A few of my friends \& i would talk to the people around us \& when enough (more than half a dozen)would agree to participate in one, we would tell one side to start \& then we would stand \& wait for the people next to us to go. By starting it we become the middle of the initial group so we call where it starts by telling who to stand first \& then we follow \& so on. 
\item[*]{\bf Please, select your geographical location.}
\\
North America 
\end{itemize}
\smallskip
\item[] {\bf \uline{(59) Apr 16, 2005}}\begin{itemize}
\item[*]{\bf Have you ever taken part in a Mexican wave?}
\\
 YES 
\item[*]{\bf If yes, did you follow your neighbours or you considered rather the motion of the wave as a whole? Please, explain below.}
\\
i dont understand the question 
\item[*]{\bf Does the wave have a preferred direction? (Clockwise or counter-clockwise?)}
\\
i think maybe counter clockwise because thats all ive ever seen 
\item[*]{\bf Have you ever seen two waves initiated by a single source? (i.e., the wave leaving in both directions) If yes, please give details, e.g., when and where.}
\\
no 
\item[*]{\bf Have you ever taken part in initiating a wave? What did you do? How many of you were needed? Did you try to influence the direction of the wave from the very beginning?}
\\
no 
\item[*]{\bf Please, select your geographical location.}
\\
North America 
\end{itemize}
\smallskip
\item[] {\bf \uline{(60) Apr 18, 2005}}\begin{itemize}
\item[*]{\bf Have you ever taken part in a Mexican wave?}
\\
 YES 
\item[*]{\bf If yes, did you follow your neighbours or you considered rather the motion of the wave as a whole? Please, explain below.}
\\
i generally tried to match the motion of the wave as a whole, attempting to stand when the wave was peaking in my section 
\item[*]{\bf Does the wave have a preferred direction? (Clockwise or counter-clockwise?)}
\\
counter-clockwise 
\item[*]{\bf Have you ever seen two waves initiated by a single source? (i.e., the wave leaving in both directions) If yes, please give details, e.g., when and where.}
\\
no 
\item[*]{\bf Have you ever taken part in initiating a wave? What did you do? How many of you were needed? Did you try to influence the direction of the wave from the very beginning?}
\\
yes. It takes either a couple of very enthusiastice people or up to six people willing to help. I generally plant helpers in the next sections of the stadium to help the wave continue. I have always tried to start waves in the counter-clockwise direction by running with uplifted arms in the direction i wish the wave to proceed. 
\item[*]{\bf Please, select your geographical location.}
\\
North America 
\end{itemize}

\subsubsection*{Answers 61 to 70 in the online survey}
\medskip
\smallskip
\item[] {\bf \uline{(61) May 8, 2005}}\begin{itemize}
\item[*]{\bf Have you ever taken part in a Mexican wave?}
\\
 YES 
\item[*]{\bf If yes, did you follow your neighbours or you considered rather the motion of the wave as a whole? Please, explain below.}
\\
as a whole 
\item[*]{\bf Does the wave have a preferred direction? (Clockwise or counter-clockwise?)}
\\
clockwise 
\item[*]{\bf Have you ever seen two waves initiated by a single source? (i.e., the wave leaving in both directions) If yes, please give details, e.g., when and where.}
\\
yes 
\item[*]{\bf Have you ever taken part in initiating a wave? What did you do? How many of you were needed? Did you try to influence the direction of the wave from the very beginning?}
\\
yes 
\item[*]{\bf Please, select your geographical location.}
\\
Europe 
\end{itemize}
\smallskip
\item[] {\bf \uline{(62) May 19, 2005}}\begin{itemize}
\item[*]{\bf Have you ever taken part in a Mexican wave?}
\\
 YES 
\item[*]{\bf If yes, did you follow your neighbours or you considered rather the motion of the wave as a whole? Please, explain below.}
\\
both - but not at the same time...NB not much thought involved - instinctive response. 
\item[*]{\bf Does the wave have a preferred direction? (Clockwise or counter-clockwise?)}
\\
clockwise 
\item[*]{\bf Have you ever seen two waves initiated by a single source? (i.e., the wave leaving in both directions) If yes, please give details, e.g., when and where.}
\\
At Queenstown Oval, Barbados. Spring 2004 Just before England Cricket team clinched the series vs. West Indies 
\item[*]{\bf Have you ever taken part in initiating a wave? What did you do? How many of you were needed? Did you try to influence the direction of the wave from the very beginning?}
\\
Again, not much thought involved. The important sensation is local response. After that theres a certain excitement in seeing the pattern spread. Much the same is true of community singing in Football and other crowds. 
\item[*]{\bf Please, select your geographical location.}
\\
Europe 
\end{itemize}
\smallskip
\item[] {\bf \uline{(63) Jun 6, 2005}}\begin{itemize}
\item[*]{\bf Have you ever taken part in a Mexican wave?}
\\
 YES 
\item[*]{\bf If yes, did you follow your neighbours or you considered rather the motion of the wave as a whole? Please, explain below.}
\\
dont understand question 
\item[*]{\bf Does the wave have a preferred direction? (Clockwise or counter-clockwise?)}
\\
no 
\item[*]{\bf Have you ever seen two waves initiated by a single source? (i.e., the wave leaving in both directions) If yes, please give details, e.g., when and where.}
\\
yes at various sporting events 
\item[*]{\bf Have you ever taken part in initiating a wave? What did you do? How many of you were needed? Did you try to influence the direction of the wave from the very beginning?}
\\
no 
\item[*]{\bf Please, select your geographical location.}
\\
North America 
\end{itemize}
\smallskip
\item[] {\bf \uline{(64) Jun 16, 2005}}\begin{itemize}
\item[*]{\bf Have you ever taken part in a Mexican wave?}
\\
 YES 
\item[*]{\bf If yes, did you follow your neighbours or you considered rather the motion of the wave as a whole? Please, explain below.}
\\
considered the motion of the wave as a whole 
\item[*]{\bf Does the wave have a preferred direction? (Clockwise or counter-clockwise?)}
\\
counter-clockwise 
\item[*]{\bf Have you ever seen two waves initiated by a single source? (i.e., the wave leaving in both directions) If yes, please give details, e.g., when and where.}
\\
No 
\item[*]{\bf Have you ever taken part in initiating a wave? What did you do? How many of you were needed? Did you try to influence the direction of the wave from the very beginning?}
\\
Yes. Scream to the crowd. Needed no more then 10.Due to teh nature of the stadum and the location we were, we tried to make teh ware go counter-clockwise. 
\item[*]{\bf Please, select your geographical location.}
\\
North America 
\end{itemize}
\smallskip
\item[] {\bf \uline{(65) Jun 22, 2005}}\begin{itemize}
\item[*]{\bf Have you ever taken part in a Mexican wave?}
\\
 YES 
\item[*]{\bf If yes, did you follow your neighbours or you considered rather the motion of the wave as a whole? Please, explain below.}
\\
The motion of the wave as a whole, i think me and my neighbour jumped up simultanously... 
\item[*]{\bf Does the wave have a preferred direction? (Clockwise or counter-clockwise?)}
\\
Clockwise 
\item[*]{\bf Have you ever seen two waves initiated by a single source? (i.e., the wave leaving in both directions) If yes, please give details, e.g., when and where.}
\\
no 
\item[*]{\bf Have you ever taken part in initiating a wave? What did you do? How many of you were needed? Did you try to influence the direction of the wave from the very beginning?}
\\
Yes. At the Olympiastation Berlin. This Statium has a tribune were no spectators can sit, so the wave could only leave in clockwise direction... We communicated with the spectators on the upper tribune and then startet to count loud backwards, jumped up and the wave was running... 
\item[*]{\bf Please, select your geographical location.}
\\
Europe 
\end{itemize}
\smallskip
\item[] {\bf \uline{(66) Jun 25, 2005}}\begin{itemize}
\item[*]{\bf Have you ever taken part in a Mexican wave?}
\\
 YES 
\item[*]{\bf If yes, did you follow your neighbours or you considered rather the motion of the wave as a whole? Please, explain below.}
\\
I simply followed those around me during the Waves first passed. On its subsequent passes (approximately 4 in all) I anticipated its motion and acted when the time seemed right. 
\item[*]{\bf Does the wave have a preferred direction? (Clockwise or counter-clockwise?)}
\\
Counter-clockwise 
\item[*]{\bf Have you ever seen two waves initiated by a single source? (i.e., the wave leaving in both directions) If yes, please give details, e.g., when and where.}
\\
No. 
\item[*]{\bf Have you ever taken part in initiating a wave? What did you do? How many of you were needed? Did you try to influence the direction of the wave from the very beginning?}
\\
No. I have never initiated a Wave. 
\item[*]{\bf Please, select your geographical location.}
\\
North America 
\end{itemize}
\smallskip
\item[] {\bf \uline{(67) Jun 29, 2005}}\begin{itemize}
\item[*]{\bf Have you ever taken part in a Mexican wave?}
\\
 YES 
\item[*]{\bf If yes, did you follow your neighbours or you considered rather the motion of the wave as a whole? Please, explain below.}
\\
I followed the motion of the wave and timed it. I dont trust my neighbors to get it right! 
\item[*]{\bf Does the wave have a preferred direction? (Clockwise or counter-clockwise?)}
\\
At Michigan Stadium in Ann Arbor, it goes counter-clockwise. 
\item[*]{\bf Have you ever seen two waves initiated by a single source? (i.e., the wave leaving in both directions) If yes, please give details, e.g., when and where.}
\\
At Michigan Stadium, there is an entire sequence that occurs....first a few rotations counter-clockwise, then it speeds up, then slows down, switches to clockwise, then splits and two waves go simultaneously in two different directions. its amazing, really. 
\item[*]{\bf Have you ever taken part in initiating a wave? What did you do? How many of you were needed? Did you try to influence the direction of the wave from the very beginning?}
\\
No. 
\item[*]{\bf Please, select your geographical location.}
\\
North America 
\end{itemize}
\smallskip
\item[] {\bf \uline{(68) Jul 6, 2005}}\begin{itemize}
\item[*]{\bf Have you ever taken part in a Mexican wave?}
\\
 YES 
\item[*]{\bf If yes, did you follow your neighbours or you considered rather the motion of the wave as a whole? Please, explain below.}
\\
followed neighbors 
\item[*]{\bf Does the wave have a preferred direction? (Clockwise or counter-clockwise?)}
\\
Counter Clock-wise 
\item[*]{\bf Have you ever seen two waves initiated by a single source? (i.e., the wave leaving in both directions) If yes, please give details, e.g., when and where.}
\\
Ive ever seen it initiated by the same group at many Florida State University football games. They get two waves going in opposite directions, however I do not believe they were started simultaniously. 
\item[*]{\bf Have you ever taken part in initiating a wave? What did you do? How many of you were needed? Did you try to influence the direction of the wave from the very beginning?}
\\
Nope, Ive always been a follower. 
\item[*]{\bf Please, select your geographical location.}
\\
North America 
\end{itemize}
\smallskip
\item[] {\bf \uline{(69) Jul 6, 2005}}\begin{itemize}
\item[*]{\bf Have you ever taken part in a Mexican wave?}
\\
 YES 
\item[*]{\bf If yes, did you follow your neighbours or you considered rather the motion of the wave as a whole? Please, explain below.}
\\
I generally anticipate the wave coming, so Im not at all dependent on my immediate neighbors. 
\item[*]{\bf Does the wave have a preferred direction? (Clockwise or counter-clockwise?)}
\\
My experience is clockwise, but I was at a show recently where it was not a stadium and the wave started left to right (from my left to my right) and then the people on the right of the audience triggered a reciprocal wave in the opposite direction 
\item[*]{\bf Have you ever seen two waves initiated by a single source? (i.e., the wave leaving in both directions) If yes, please give details, e.g., when and where.}
\\
no 
\item[*]{\bf Have you ever taken part in initiating a wave? What did you do? How many of you were needed? Did you try to influence the direction of the wave from the very beginning?}
\\
Ive started a couple of waves. Typically I start with only a handful of people and then recruit those folks sitting near us. Not consciously, but both times, I recurited people sitting to our left, therefore creating a clockwise wave. 
\item[*]{\bf Please, select your geographical location.}
\\
North America 
\end{itemize}
\smallskip
\item[] {\bf \uline{(70) Jul 16, 2005}}\begin{itemize}
\item[*]{\bf Have you ever taken part in a Mexican wave?}
\\
 YES 
\item[*]{\bf If yes, did you follow your neighbours or you considered rather the motion of the wave as a whole? Please, explain below.}
\\
i followed the prson beside me 
\item[*]{\bf Does the wave have a preferred direction? (Clockwise or counter-clockwise?)}
\\
not really, it usually starts from one side and then goes backward 
\item[*]{\bf Have you ever seen two waves initiated by a single source? (i.e., the wave leaving in both directions) If yes, please give details, e.g., when and where.}
\\
no, i havent 
\item[*]{\bf Have you ever taken part in initiating a wave? What did you do? How many of you were needed? Did you try to influence the direction of the wave from the very beginning?}
\\
We didnt really do much, we just yelled Una Ola!! una Ola!!!! (a wave!) and got up, and everyone followed. We indicated as we yelled which way to go, and it came back to us as it finished on one side. It only took about five of us to start it. 
\item[*]{\bf Please, select your geographical location.}
\\
North America 
\end{itemize}

\subsubsection*{Answers 71 to 75 in the online survey}
\medskip
\smallskip
\item[] {\bf \uline{(71) Jul 19, 2005}}\begin{itemize}
\item[*]{\bf Have you ever taken part in a Mexican wave?}
\\
 YES 
\item[*]{\bf If yes, did you follow your neighbours or you considered rather the motion of the wave as a whole? Please, explain below.}
\\
Neighbour. THe moment he or she gets up, I start to. My imperfect reaction speed creates enough delay to continue to wave effect. 
\item[*]{\bf Does the wave have a preferred direction? (Clockwise or counter-clockwise?)}
\\
Clockwise 
\item[*]{\bf Have you ever seen two waves initiated by a single source? (i.e., the wave leaving in both directions) If yes, please give details, e.g., when and where.}
\\
No 
\item[*]{\bf Have you ever taken part in initiating a wave? What did you do? How many of you were needed? Did you try to influence the direction of the wave from the very beginning?}
\\
No 
\item[*]{\bf Please, select your geographical location.}
\\
North America 
\end{itemize}
\smallskip
\item[] {\bf \uline{(72) Aug 5, 2005}}\begin{itemize}
\item[*]{\bf Have you ever taken part in a Mexican wave?}
\\
 YES 
\item[*]{\bf If yes, did you follow your neighbours or you considered rather the motion of the wave as a whole? Please, explain below.}
\\
The motion of the wave as a whole. In a baseball game, the focus of almost the entire crowd is on the wave, not on the game. You cant help but follow the wave and anticipate its arrival. One of the things I noticed about your simulation is that it will either die out immediately or propagate forever. My experience is that sometimes a wave will travel halfway around the stadium before dying out completely. In terms of your model, the activation threshold varies in a systematic manner (Usually the cheap seats have the easiest threshold, while the more expensive seats are harder to activate) 
\item[*]{\bf Does the wave have a preferred direction? (Clockwise or counter-clockwise?)}
\\
It depends entirely on the person who starts it. I have seen waves go in both directions in the same stadium (at different events) 
\item[*]{\bf Have you ever seen two waves initiated by a single source? (i.e., the wave leaving in both directions) If yes, please give details, e.g., when and where.}
\\
No, since the person starting the wave indicates a preffered direction and either the wave follows him, or doesnt start at all. 
\item[*]{\bf Have you ever taken part in initiating a wave? What did you do? How many of you were needed? Did you try to influence the direction of the wave from the very beginning?}
\\
I have not started a wave myself, but I have been very close to and observant of those who did in several cases. It is usually just one or two people near the front row of a section. He or they will stand up, turn around and try to get the attention of the section he is sitting in, tell them they are going to do a wave, then count to 3. At 3 he raises his arms up and to the side, and if he has room, he runs in the direction he wants the wave to go. It usually takes several tries. If conditions are right, his section will react very well, but the wave will die out after travelling only a few sections. People farther away just arent paying attention and dont notice the wave. After several attempts, the sections farther and farther away start paying attention and the wave goes further before dying out. If the wave can make it through the most heavily seated section with significant energy, it will probably make it all the way around and be self-sustaining. Sometimes it is the cheerleaders (in american college football games) who start it. They will run on the sidelines right near the front row with a big flag and start it. They usually take fewer tries to start a wave, but they are still rarely successful on the first try. I have seen waves in baseball stadiums and an american football stadium. Baseball stadiums almost always have fewer rows of seats in the outfield, and the most around home plate. Waves are usually started by someone in the outfield seats and sent towards home plate in whichever direction seems shortest. My local football stadium has many fewer rows at one end zone (a U shape). Again, waves are usually sent in the direction of more seats. 
\item[*]{\bf Please, select your geographical location.}
\\
North America 
\end{itemize}
\smallskip
\item[] {\bf \uline{(73) Aug 11, 2005}}\begin{itemize}
\item[*]{\bf Have you ever taken part in a Mexican wave?}
\\
 YES 
\item[*]{\bf If yes, did you follow your neighbours or you considered rather the motion of the wave as a whole? Please, explain below.}
\\
I think its a function of the two. As I see it, in a normal stadium you are able to see very well the wave approach from afar, but it is more difficult to see where the wave is as it quickly approaches you. I considered the motion of the wave as a whole for timing of when it would approximatly get to me and then I used my neighbours as the key signal as to when I should stand. 
\item[*]{\bf Does the wave have a preferred direction? (Clockwise or counter-clockwise?)}
\\
I think it depends on the stadium and the sparking group. My last wave experience it went counter-clockwise. Although i could not see where it started, I think the sparking group was in the top level of left field. There is a gap between this area and right field. The wave must go counter-clockwise if it is started there. It would be interesting to include a description of the stadium and the location of the waves starting point in relationship to direction. 
\item[*]{\bf Have you ever seen two waves initiated by a single source? (i.e., the wave leaving in both directions) If yes, please give details, e.g., when and where.}
\\
No. I also think it would shorten the number of revolutions due to the law of deminishing returns. 
\item[*]{\bf Have you ever taken part in initiating a wave? What did you do? How many of you were needed? Did you try to influence the direction of the wave from the very beginning?}
\\
--
\item[*]{\bf Please, select your geographical location.}
\\
North America 
\end{itemize}
\smallskip
\item[] {\bf \uline{(74) Aug 17, 2005}}\begin{itemize}
\item[*]{\bf Have you ever taken part in a Mexican wave?}
\\
 YES 
\item[*]{\bf If yes, did you follow your neighbours or you considered rather the motion of the wave as a whole? Please, explain below.}
\\
both. It moves very quick when it gets near you, so you must plan ahead. 
\item[*]{\bf Does the wave have a preferred direction? (Clockwise or counter-clockwise?)}
\\
always has been clockwise. 
\item[*]{\bf Have you ever seen two waves initiated by a single source? (i.e., the wave leaving in both directions) If yes, please give details, e.g., when and where.}
\\
no, never. 
\item[*]{\bf Have you ever taken part in initiating a wave? What did you do? How many of you were needed? Did you try to influence the direction of the wave from the very beginning?}
\\
no, Ive always participated in an existing wave. actually, Ive never even directly observed one start. 
\item[*]{\bf Please, select your geographical location.}
\\
North America 
\end{itemize}
\smallskip
\item[] {\bf \uline{(75) Aug 18, 2005}}\begin{itemize}
\item[*]{\bf Have you ever taken part in a Mexican wave?}
\\
 YES 
\item[*]{\bf If yes, did you follow your neighbours or you considered rather the motion of the wave as a whole? Please, explain below.}
\\
Both--watched wave approach, but waited until immediate neighbors started. 
\item[*]{\bf Does the wave have a preferred direction? (Clockwise or counter-clockwise?)}
\\
Clockwise 
\item[*]{\bf Have you ever seen two waves initiated by a single source? (i.e., the wave leaving in both directions) If yes, please give details, e.g., when and where.}
\\
No 
\item[*]{\bf Have you ever taken part in initiating a wave? What did you do? How many of you were needed? Did you try to influence the direction of the wave from the very beginning?}
\\
No 
\item[*]{\bf Please, select your geographical location.}
\\
North America 
\end{itemize}
\end{itemize}
}

\end{document}